\documentclass[11pt,a4paper]{article}
\pdfoutput=1
\usepackage{cancel}
\usepackage{enumitem}
\usepackage{etex}
\usepackage{amsthm}
\usepackage{geometry}
\usepackage{dcolumn}
\usepackage{epsf}
\usepackage{mathrsfs}
\usepackage{multirow}
\usepackage{booktabs}
\usepackage{tabularx}
\usepackage{array}
\usepackage{slashed}
\usepackage{float}
\usepackage{leftidx}
\usepackage{setspace}
\usepackage{verbatim}
\usepackage{adjustbox}
\usepackage{tikz}
\usepackage[caption=false]{subfig}
\usepackage{arydshln}
\usepackage{jheppub}
\usepackage{shuffle}
\usepackage{amsmath}
\usepackage{cases}

\numberwithin{equation}{section}
\usetikzlibrary{arrows,decorations.markings,shapes.arrows,patterns,positioning}

\newcommand{\bea}{\begin{eqnarray}}
\newcommand{\eea}{\end{eqnarray}}
\newcommand{\bean}{\begin{eqnarray*}}
\newcommand{\eean}{\end{eqnarray*}}
\newcommand{\nn}{\nonumber\\}
\newcommand{\Sl}{\sum\limits}

\def\W #1{\widetilde{#1}}

\def\Label#1{\label{#1}%
  \smash{\hbox to0pt{\raise1ex\hbox{\tiny[#1]}\hss}}}
\def\Label#1{\label{#1}}
\renewcommand{\eqref}[1]{eq.~(\ref{#1})}
\newcommand{\figref}[1]{Fig.~\ref{#1}}

\newcommand{\secref}[1]{section~\ref{#1}}
\newcommand{\appref}[1]{appendix~\ref{#1}}

\def\braket#1{\left\langle #1 \right\rangle}

\def\vev{\braket}

\def\bvev#1{\left[ #1 \right]}
\def\Spaa{\vev}
\def\Spbb{\bvev}

\def\Sl{\sum\limits}

\newcommand{\ctobedelete}[1]{}

\allowdisplaybreaks

\title{A note on multi-trace EYM amplitudes in four dimensions}

\author[a]{Chongsi Xie} \author[a,b]{Yi-Jian Du\footnote{Corresponding author}}

\affiliation[a]{Department of Physics, School of Physics and Technology,
Wuhan University, \\
No.299 Bayi Road, Wuhan 430072, China}
\affiliation[b]{Hubei Key Laboratory of Nuclear Solid Physics, School of Physics and Technology, Wuhan University,\\
No.299 Bayi Road, Wuhan 430072, China}
\emailAdd{chongsi.xie@whu.edu.cn} \emailAdd{yijian.du@whu.edu.cn}

\date{\today}
\abstract{In four dimensions, a tree-level double-trace Einstein-Yang-Mills (EYM) amplitude with two negative-helicity gluons (the $(g^-,g^-)$-configuration) satisfies a symmetric spanning forest formula, which was derived from the graphic expansion rule. On another hand, in the framework of Cachazo-He-Yuan (CHY) formula, the maximally-helicity-violating (MHV) amplitudes are supported by the MHV  solution of scattering equations.  The relationship between the symmetric formula for double-trace amplitudes, and the MHV sector of Cachazo-He-Yuan (CHY) formula in four dimensions is still not clear. In this note, we promote a series of transformations of the spanning forests in four dimensions and then show a systematic way for decomposing the MHV sector of the CHY formula of double-trace EYM amplitudes. Along this line, the symmetric formula of double-trace MHV amplitudes is directly obtained by the MHV sector of CHY formula. We then prove that  EYM amplitude with an arbitrary total number of negative-helicity particles (gravitons and gluons) has to vanish when the number of negative- (or positive-) helicity gluons is less than the number of traces.}


\begin{document}
\maketitle \flushbottom

\section{Introduction}

The expansion of Einstein-Yang-Mills (EYM)\footnote{In this paper, B-field and dilaton are also involved in the theory.} amplitudes states that tree-level EYM amplitudes can be expanded in terms of color-ordered Yang-Mills (YM) amplitudes. This expansion was first observed for single-trace amplitudes with one graviton \cite{Stieberger:2016lng}, then extended to amplitudes with a few gravitons and gluon traces \cite{Nandan:2016pya,Schlotterer:2016cxa}.
General patterns of the expansions based on a recursive expression were established for not only single-trace amplitudes \cite{Fu:2017uzt,Chiodaroli:2017ngp,Teng:2017tbo} but also multi-trace ones \cite{Du:2017gnh}. These expansion formulas  provided a new angle for understanding several related problems, including: constructing local Bern-Carrasco-Johansson (BCJ) \cite{Bern:2008qj} numerators \cite{Fu:2017uzt,Du:2017kpo,Du:2018khm}, inducing new amplitude relations by gauge symmetries \cite{Fu:2017uzt,Du:2017kpo,Du:2018khm,Hou:2018bwm,Du:2019vzf}, providing an off-shell approach to BCJ duality (and amplitude relations) \cite{Wu:2021exa,Du:2022vsw} and evaluating EYM amplitudes in four dimensions \cite{Tian:2021dzf}. Several discussions at loop amplitudes can also be found (see \cite{Porkert:2022efy,Zhou:2022djx,Faller:2018vdz} for example).

Among the above progresses on the expansion of EYM amplitudes, a symmetric formula of maximally-helicity-violating (MHV) double-trace EYM amplitudes in four dimensions was proposed  \cite{Tian:2021dzf}. In this formula, the two gluon traces are given by two Parke-Taylor \cite{Parke:1986gb} factors, while gravitons are included by spanning forests that are rooted at gluons. A key feature of this formula is that either the two traces or the gravitons are arranged in an equal footing. Such symmetric pattern was earlier discovered for the tree-level gravity (GR) amplitudes \cite{Nguyen:2009jk,Hodges:2012ym,Feng:2012sy}, the tree-level single-trace EYM amplitudes \cite{Du:2016wkt} as well as the double-trace pure gluon amplitudes in EYM theory \cite{Cachazo:2014nsa}, with the MHV configurations.

Apart from the line of EYM expansion, the famous Cachazo-He-Yuan (CHY) \cite{Cachazo:2013gna,Cachazo:2013hca,Cachazo:2013iea,Cachazo:2014nsa,Cachazo:2014xea} formula provides another powerful approach to the calculations in four dimensions. By CHY formula, one can substitute the MHV solution (i.e.  the solution supporting the MHV amplitudes) of the scattering equations (SE) into the CHY integrand, which involves a Pfaffian  and a reduced Pfaffian, straightforwardly and then expand the integrand in a proper way to obtain an expression of MHV amplitudes. This approach verifies the fact that the MHV solution supports the MHV amplitudes. Along this line, the tree-level MHV amplitudes of YM, GR and the single-trace MHV EYM amplitudes have been already evaluated \cite{Du:2016blz,Du:2016wkt}. Nevertheless, there are still gaps between the symmetric formula proposed in \cite{Tian:2021dzf} and the direct evaluation of the CHY formula for double-trace MHV amplitudes. First, the two traces in the CHY formula do not stand in an equal status, in the sense that the Pfaffian part in CHY formula only involves gluons from one trace. Thus the relationship between the MHV sector of CHY formula and the symmetric formula in  \cite{Tian:2021dzf} is not transparent.  Second, in \cite{Weinzierl:2014vwa,Du:2016fwe}, it was explicitly shown that the reduced Pfaffian with two negative-helicity particles is supported by the MHV solution to SE\footnote{Studies of solutions to SE from distinct aspects can be found in the work \cite{Roberts:1972abc,Fairlie:1972abc,Fairlie:2008dg,Cachazo:2013hca,Cachazo:2013gna,Monteiro:2013rya,Geyer:2014fka,Mason:2013sva,Weinzierl:2014vwa,Dolan:2015iln,Cardona:2015ouc,Cachazo:2016sdc,He:2016vfi,Du:2016fwe}.} by substituting the MHV solution into the reduced Pfaffian, but there are vanishing configurations \cite{Cachazo:2014xea,Tian:2021dzf} of multi-trace amplitudes with two negative-helicity particles.  More generally, the vanishing configurations for amplitudes with an arbitrary number of negative-helicity particles for GR, YM, single-trace EYM amplitudes and  pure gluon multi-trace EYM were already understood in a way via considering the rank of  a 
discriminant matrix (see \cite{Cachazo:2014xea} for the pure-gluon multi-trace case  and \cite{Du:2016fwe} for the single-trace EYM cases with gravitons). However, these arguments have not been extended to multi-trace cases with an arbitrary number of gravitons yet. In this note, we fill these two gaps. By breaking the symmetry between the two traces and proposing a new formula of double-trace MHV amplitudes, we find the connection between the MHV sector of CHY formula and the symmetric formula of double-trace amplitudes \cite{Tian:2021dzf}. We further prove that amplitudes with an arbitrary number of negative- (positive-) helicity particles  have to vanish when the number of negative- (and/or positive-) helicity gluons is less than the number of gluon traces. This is a complement of the discussion on the vanishing configurations \cite{Cachazo:2014xea,Du:2016fwe}.


The structure of this note is arranged as follows. In \secref{SE:NewFormula}, we introduce several helpful transformations of the spanning forests and then prove  a new formula of double-trace MHV amplitude with two negative-helicity gluons, by breaking the symmetry between the two traces. In \secref{SE:CHYApproach}, we derive the formula proposed in  \secref{SE:NewFormula} along the other line, i.e. the CHY approach, by substituting the MHV solution to SE into Pfaffian.
The vanishing configurations for an arbitrary number of negative-helicity particles are studied in \secref{SE:VanishingCoufigurations}. Conclusions of this note are provided in \secref{SE:Conclusions}. A one-graviton example for the proof in \secref{Sec:GenProof} and the vanishing condition of a determinant introduced in \secref{SE:VanishingCoufigurations} are provided in the appendix.

\begin{figure}
\centering
\includegraphics[width=0.65\textwidth]{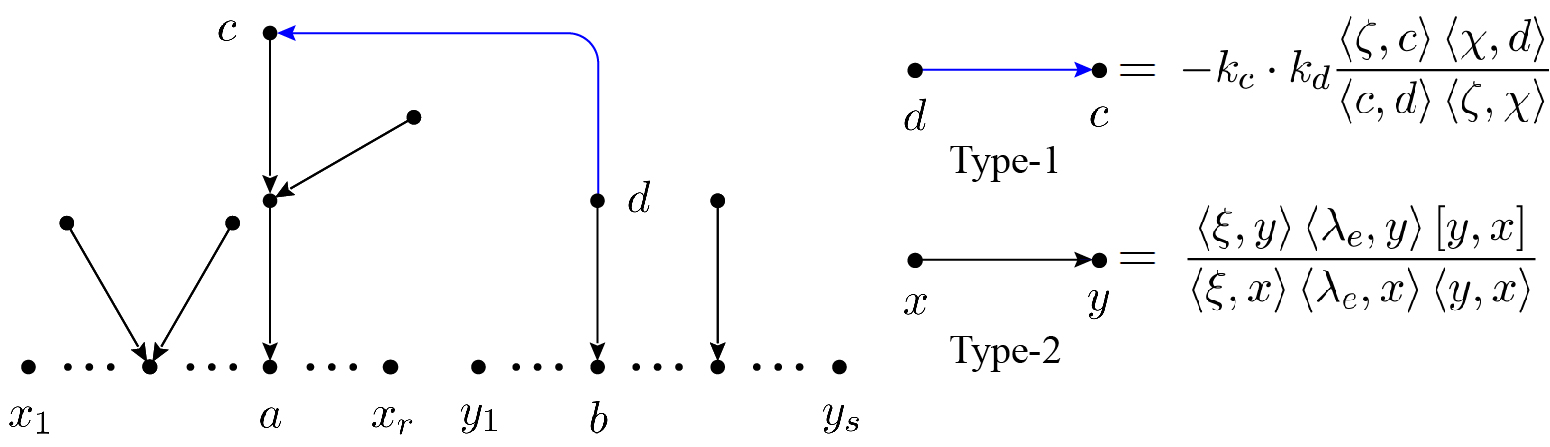}
\caption{A typical spanning forest for double-trace $(g^-,g^-)$-amplitude, which includes a bridge connecting two traces by a type-1 edge and several type-2 edges.}
\label{Fig:Spanningforest}
\end{figure}
%
\section{A new formula of double-trace MHV amplitudes with $(g^-,g^-)$-configuration}\label{SE:NewFormula}

In EYM theory, a tree-level double-trace MHV amplitude $A^{(g^-_i,g^-_j)}(x_1,\ldots, x_r|y_1,\ldots,y_s\Vert \mathsf{H})$ ($g_i$, $g_j$ are the two negative-helicity gluons) with two gluon traces  $\pmb{1}=\{x_1,\dots,x_r\}$, $\pmb{2}=\{y_1,\dots,y_s\}$ and the graviton set $\mathsf{H}$  satisfies the following symmetric spanning forest formula \cite{Tian:2021dzf}
\bea
&&A^{(g^-_i,g^-_j)}(x_1,\ldots, x_r|y_1,\ldots,y_s\Vert \mathsf{H})\nn
&=&\frac{\Spaa{g_i,g_j}^4}{\left(x_1,\dots,x_r\right)\left(y_1,\dots,y_s\right)}\Bigg[\,\Sl_{\mathsf{H}\to \mathsf{H}_A, \mathsf{H}_B}\,\Sl_{{\mathcal{B}}}\,\left(-k_c\cdot k_d\right)\frac{\Spaa{\zeta,c}\Spaa{\chi,d}}{\Spaa{c,d}\Spaa{\zeta,\chi}}\,\prod\limits_{{e(x,y)\in\,\mathcal{E}_1(\mathcal{B})}}\frac{\Spaa{\xi,y}\Spaa{\zeta,y}\Spbb{y,x}}{\Spaa{\xi,x}\Spaa{\zeta,x}\Spaa{y,x}}\nn
&&\times\prod\limits_{{e(x,y)\in\,\mathcal{E}_2(\mathcal{B})}}\frac{\Spaa{\xi,y}\Spaa{\chi,y}\Spbb{y,x}}{\Spaa{\xi,x}\Spaa{\chi,x}\Spaa{y,x}}\,\Sl_{\mathcal{G}}\,\prod\limits_{{e(x,y)\in\,\mathcal{E}(\mathcal{G})}}\frac{\Spaa{\xi,y}\Spaa{\eta,y}\Spbb{y,x}}{\Spaa{\xi,x}\Spaa{\eta,x}\Spaa{y,x}}\Bigg] ,\Label{Eq:DoubleTraceMHV1}
\eea
which is written by spinor products \cite{Xu:1986xb}. On the RHS, the prefactor $\frac{\Spaa{g_i,g_j}^4}{\left(x_1,\dots,x_r\right)\left(y_1,\dots,y_s\right)}$ involves two Parke-Taylor \cite{Parke:1986gb} factors which encode permutations of gluons in the traces, while the expression in the square brackets is interpreted by spanning forests where gluons and gravitons are considered as nodes. The spanning forests are generated by the following steps:
\begin{itemize}
\item {\bf Step-1}~~Split the graviton set $\mathsf{H}$ into two subsets $\mathsf{H}_A$ and $\mathsf{H}_B$.

\item  {\bf Step-2}~~Choose a pair of gluons $a\in\pmb{1}$, $b\in\pmb{2}$. Construct a path towards $a$ passing through some elements of $\mathsf{H}_A$, and a path towards $b$ passing through the remaining elements of $\mathsf{H}_A$. Connecting the starting nodes $c$ and $d$ of the two paths, one obtains a bridge $\mathcal{B}$ between the two traces.

\item  {\bf Step-3}~~For a given bridge constructed by the previous step, generate a spanning forest ${\mathcal{G}}$ which connects the gravitons in $\mathsf{H}_B$ to roots (i.e. gluons and the gravitons on the bridge) via tree structures.

\end{itemize}
A spanning forest, as shown in \figref{Fig:Spanningforest}, that is  generated by the above steps contributes the product of the factors corresponding to each edge. The edge $e(d,c)$ between the nodes $c$ and $d$ in step-2 (the blue edge in  \figref{Fig:Spanningforest}) is called \emph{type-1 edge} and is accompanied by a factor
\bea
-k_c\cdot k_d\frac{\Spaa{\zeta,c}\Spaa{\chi,d}}{\Spaa{c,d}\Spaa{\zeta,\chi}},\Label{Eq:Type-1}
\eea
while other edges $e(x,y)$ pointing from $x$ to $y$ correspond to
\bea
\frac{\Spaa{\xi,y}\Spaa{\lambda_e,y}\Spbb{y,x}}{\Spaa{\xi,x}\Spaa{\lambda_e,x}\Spaa{y,x}},\,\,\, \text{where}\,\,\, \lambda_e=\left\{\small
            \begin{array}{cc}
              \eta, & e(x,y)\in\,\mathcal{E}(\mathcal{G}) \\
               \zeta, & e(x,y)\in\,\mathcal{E}_1(\mathcal{B}) \\
              \chi, & e(x,y)\in\,\mathcal{E}_2(\mathcal{B})
            \end{array}\,\right.,
\eea
are mentioned as \emph{type-2 edges}. Here, the set of edges on the \emph{bridge} pointing to the trace $\pmb{1}$ and the trace $\pmb{2}$ are denoted by $\mathcal{E}_1(\mathcal{B})$ and $\mathcal{E}_2(\mathcal{B})$, respectively, while the set of \emph{other edges} are denoted by $\mathcal{E}(\mathcal{G})$.
The $\eta$, $\zeta$, $\chi$ and $\xi$ are arbitrarily chosen (but $\zeta\neq \chi$ to avoid the divergence in (\ref{Eq:Type-1})) reference spinors which reflect symmetries of the amplitude  \cite{Tian:2021dzf}. For convenience, we set $\eta=\chi$ in this paper. Under this choice, the gauges for edges in $\mathcal{E}(\mathcal{G})$ and $\mathcal{E}_2(\mathcal{B})$ are identical but they are distinct from those for $\mathcal{E}_1(\mathcal{B})$.  A typical forest is given by  \figref{Fig:Spanningforest}.
\begin{figure}
\centering
\includegraphics[width=0.6\textwidth]{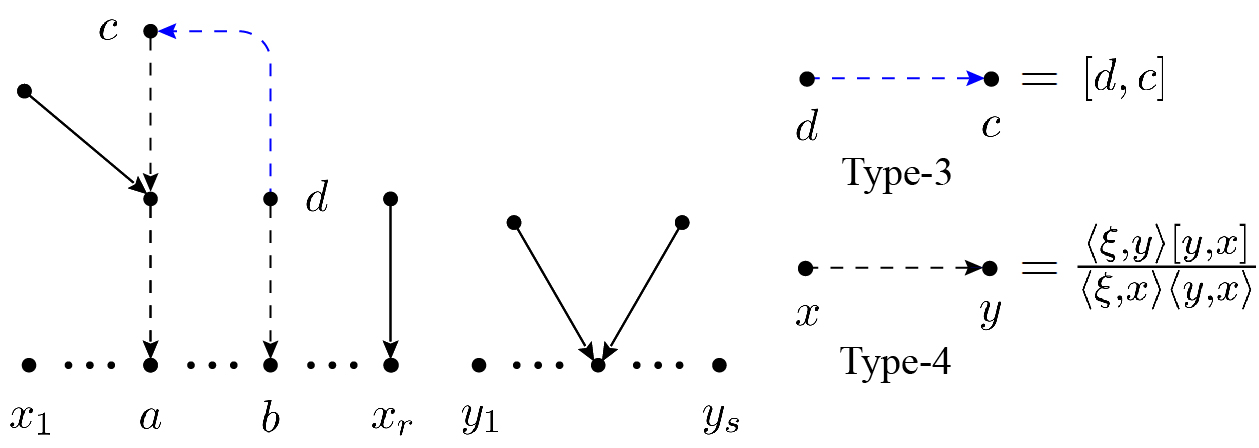}
\caption{A typical new graph for double-trace $(g^-,g^-)$-amplitude consists of two mutually disjoint parts, while the left part corresponding to the trace $\pmb{1}$ contains a bridge formed by a type-3 edge and several type-4 edges.}
\label{Fig:Graph1}
\end{figure}
%

%
%

In this section, we introduce the following new formula for double-trace MHV amplitudes
\bea
A^{(g^-_i,g^-_j)}(x_1,\ldots, x_r|y_1,\ldots,y_s\Vert \mathsf{H})&=&\frac{\Spaa{g_i,g_j}^4}{\left(x_1,\dots,x_r\right)\left(y_1,\dots,y_s\right)}\Bigg[\,\Sl_{\mathsf{H}\to \mathsf{H}_A, \mathsf{H}_B}\,\Sl_{\mathcal{B}'}\,\Spaa{a,b}\Spbb{d,c}\nn
&&\times\prod\limits_{{e(x,y)\in\,\mathcal{E}(\mathcal{B}')}}\frac{\Spaa{\xi,y}\Spbb{y,x}}{\Spaa{\xi,x}\Spaa{y,x}}\,\Sl_{\mathcal{G}}\,\prod\limits_{{e(x,y)\in\,\mathcal{E}(\mathcal{G})}}\frac{\Spaa{\xi,y}\Spaa{\eta,y}\Spbb{y,x}}{\Spaa{\xi,x}\Spaa{\eta,x}\Spaa{y,x}}\Bigg]\,\, ,
\Label{Eq:DoubleTraceMHV2}
\eea
which is equivalent to \eqref{Eq:DoubleTraceMHV1}. Being different from \eqref{Eq:DoubleTraceMHV1}, the above equation is based on spanning forests  consisting of two mutually disjoint components,  which correspond to the two gluon traces, as shown by \figref{Fig:Graph1}. In \figref{Fig:Graph1}, the two ending nodes $a$ and $b$ of the bridge $\mathcal{B}'$ now become gluons belonging to the same trace $\pmb{1}$ and the node $a$ is always nearer to the gluon $x_1\in \pmb{1}$ than the node $b$. The bridge in \figref{Fig:Graph1} is expressed by type-3 and -4 edges $e(d,c)$, $e(x,y)$ instead, which correspond to the factors  $\Spbb{d,c}$, $\frac{\Spaa{\xi,y}\Spbb{y,x}}{\Spaa{\xi,x}\Spaa{y,x}}$. Other nodes are connected to gluon traces or bridge through type-2 edges.

In the following, helpful transformations satisfied by the spanning forests are introduced, by which we prove that the formula (\ref{Eq:DoubleTraceMHV1}) where the bridge is constructed between distinct traces can be transformed into the formula (\ref{Eq:DoubleTraceMHV2}) where both ends of the bridge are moved to a single trace.

\subsection{Transformations  of spanning forests}

To show the two formulas (\ref{Eq:DoubleTraceMHV1}) and  (\ref{Eq:DoubleTraceMHV2}) are equivalent to each other, we introduce the following three transformations of spanning forests.

{\bf {Transformation-1:}}
\begin{figure}
\centering
\includegraphics[width=0.65\textwidth]{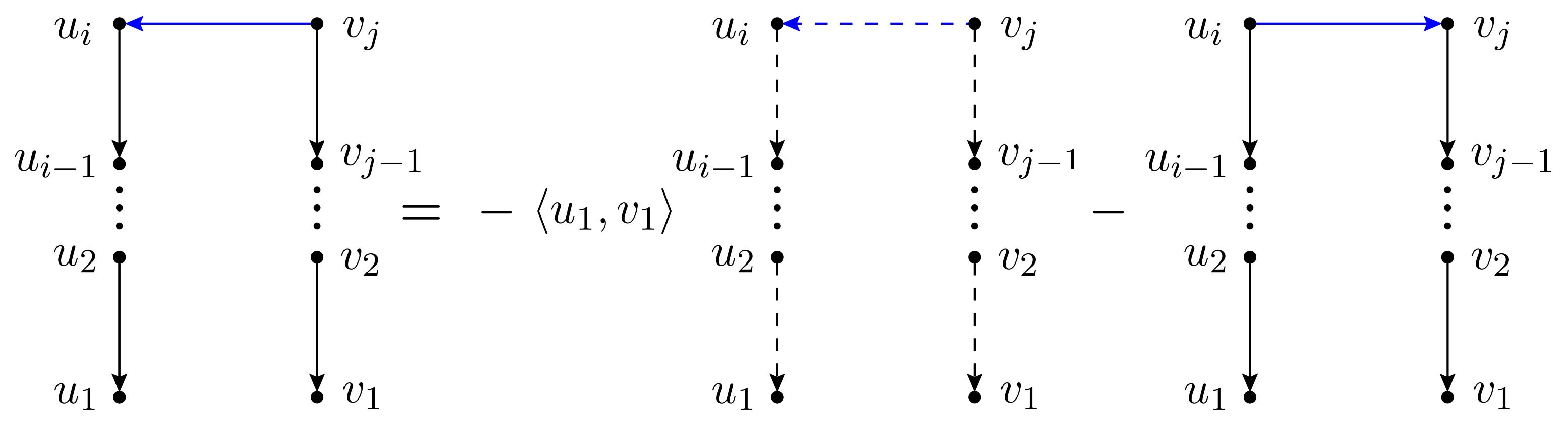}
\caption{Transformation-1 of the bridge of spanning forest (The type-2 edge in the same direction as the type-1 edge is accompanied by the reference spinor $\zeta$, while the type-2 edge in the opposite direction to the type-1 edge is accompanied by the reference spinor $\chi$.)}
\label{Fig:Property-1}
\end{figure}
When the $k_c\cdot k_d$ in (\ref{Eq:Type-1}) is expressed by spinor products, the expression corresponding to the structure on the LHS of \figref{Fig:Property-1} is given by
\bea
\Spbb{u_i,v_j}\frac{\cancel{\Spaa{\zeta,u_i}}\cancel{\Spaa{\chi,v_j}}}{\Spaa{\zeta,\chi}}\frac{\Spaa{\xi,u_1}\Spaa{\zeta,u_1}\Spbb{u_1,u_2}}{\Spaa{\xi,u_2}\cancel{\Spaa{\zeta,u_2}}\Spaa{u_1,u_2}}\frac{\Spaa{\xi,u_2}\cancel{\Spaa{\zeta,u_2}}\Spbb{u_2,u_3}}{\Spaa{\xi,u_3}\cancel{\Spaa{\zeta,u_3}}\Spaa{u_2,u_3}}\dots\frac{\Spaa{\xi,u_{i-1}}\cancel{\Spaa{\zeta,u_{i-1}}}\Spbb{u_{i-1},u_i}}{\Spaa{\xi,u_i}\cancel{\Spaa{\zeta,u_i}}\Spaa{u_{i-1},u_i}}\nn
\times\frac{\Spaa{\xi,v_1}\Spaa{\chi,v_1}\Spbb{v_1,v_2}}{\Spaa{\xi,v_2}\cancel{\Spaa{\chi,v_2}}\Spaa{v_1,v_2}}\frac{\Spaa{\xi,v_2}\cancel{\Spaa{\chi,v_2}}\Spbb{v_2,v_3}}{\Spaa{\xi,v_3}\cancel{\Spaa{\chi,v_3}}\Spaa{v_2,v_3}}\dots\frac{\Spaa{\xi,v_{j-1}}\cancel{\Spaa{\chi,v_{j-1}}}\Spbb{v_{j-1},v_j}}{\Spaa{\xi,v_j}\cancel{\Spaa{\chi,v_j}}\Spaa{v_{j-1},j_i}}\,\, .
\Label{Eq:Property-1.1}
\eea
\begin{figure}
\centering
\includegraphics[width=0.77\textwidth]{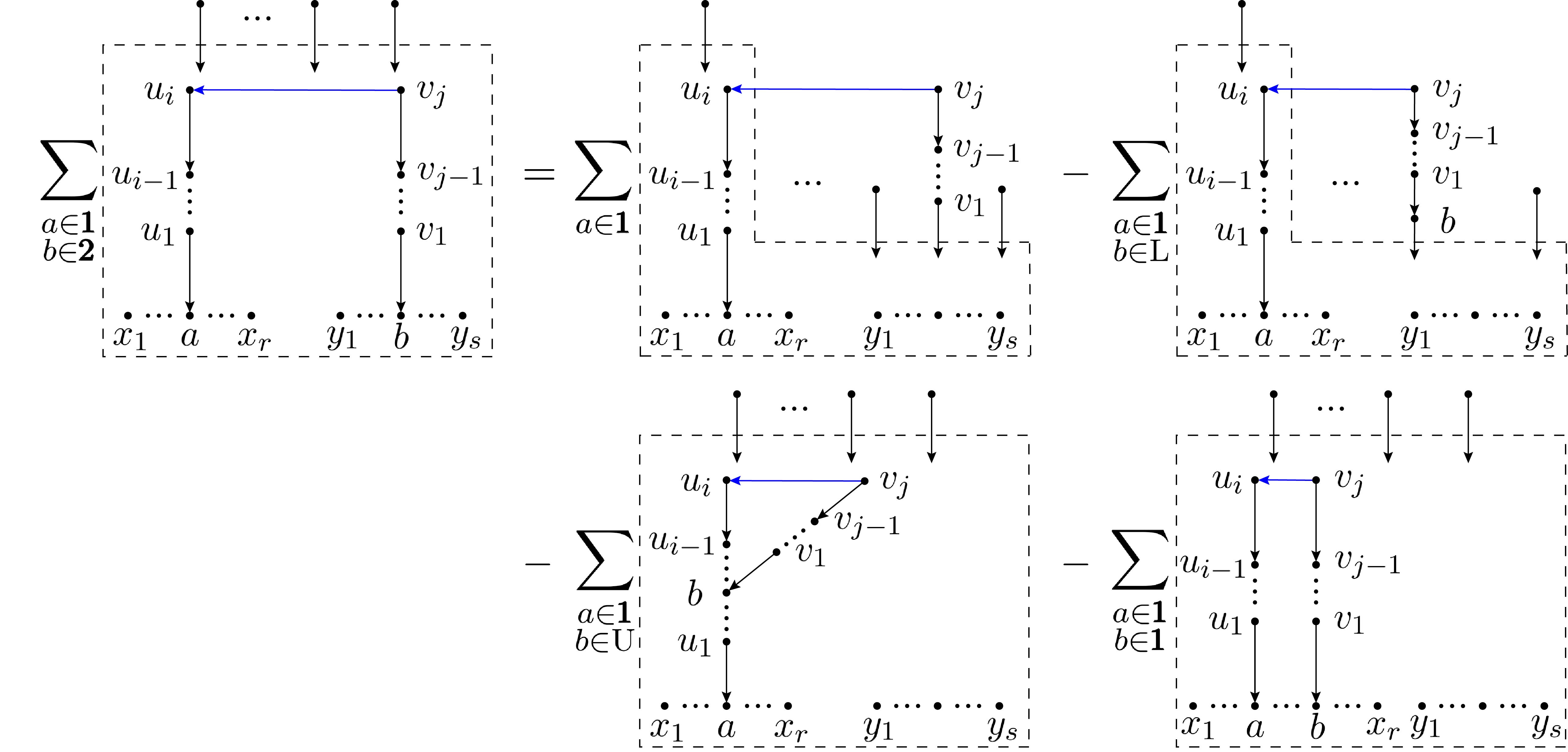}
\caption{Transformation-2 of spanning forests (The set $\{u_1, u_2\dots u_i\}$ is denoted by $U$, and the set of gravitons that are not in set $\{u_1, u_2\dots u_i, v_1, v_2, \dots v_j\}$ is denoted by $L$.)  Each graph also imply a sum over all possible spannning forests generated by nodes outside the boxed region. }
\label{Fig:Property-2}
\end{figure}
Note that the minus of the factor for the type-3 edge $e(v_j,u_i)$ is absorbed due to the antisymmetry of the spinor product $\Spbb{u_i,v_j}=-\Spbb{v_j,u_i}$. After dividing out the common factors in the numerator and denominator, the factors involving the reference spinors $\zeta$ and $\chi$ are collected as
\bea
\frac{\Spaa{\zeta,u_1}\Spaa{\chi,v_1}}{\Spaa{\zeta,\chi}}=-\Spaa{v_1,u_1}+\frac{\Spaa{\zeta,v_1}\Spaa{\chi,u_1}}{\Spaa{\zeta,\chi}}\,\, ,
\Label{Eq:Property-1.2}
\eea
where the Schouten identity $\Spaa{a,b}\Spaa{c,d}+\Spaa{a,c}\Spaa{d,b}+\Spaa{a,d}\Spaa{b,c}=0$ has been applied. When the RHS of \eqref{Eq:Property-1.2} are inserted, the expression (\ref{Eq:Property-1.1}) splits into two terms corresponding to the graphs on the RHS of  \figref{Fig:Property-1}, where the first contains the type-3 and -4 edges introduced in \eqref{Eq:DoubleTraceMHV2} and the second still contains the type-1 and -2 edges in \eqref{Eq:DoubleTraceMHV1} but the choices of gauge on both sides of the type-1 edge are exchanged.

{\bf {Transformation-2:}}
\begin{figure}
\centering
\includegraphics[width=1.0\textwidth]{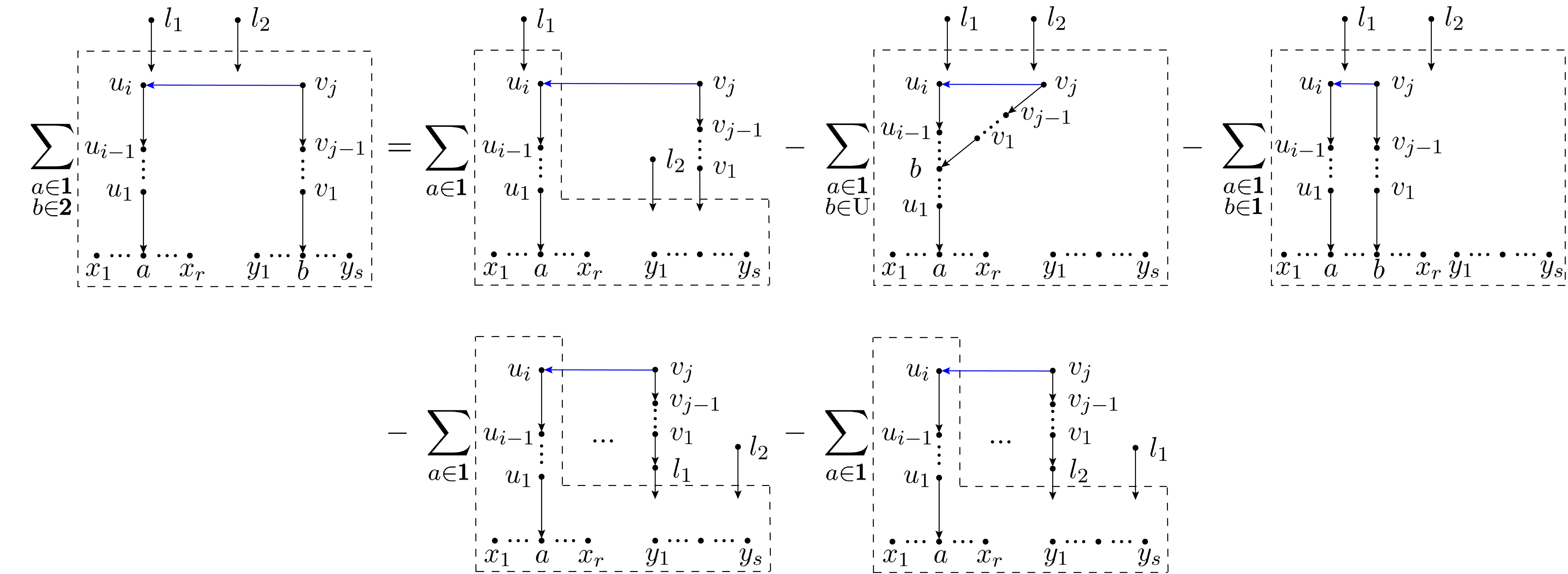}
\caption{Transformation-2 of spanning forests with two gravitons $l_1$ and $l_2$ outside the boxed region}
\label{Fig:example1}
\end{figure}
For given $a$, $b$ on the LHS of  \figref{Fig:Property-2}, all nodes in the boxed region are considered as roots. Other nodes are connected to the roots via tree structures. The summation over all possible such spanning forests are implied by the box. In the first graph on the RHS, we sum over all possible spanning forests with keeping the chain  structure $v_j\to v_{j-1}\to...\to v_1$ and then connect the node $v_j$ to $u_i$ via a type-1 edge. The contributions where $b$ (the root of the chain $v_j\to v_{j-1}\to...\to v_1$) is (i). a node outside the boxed regions, (ii). a node belonging to $\{u_1,...,u_i\}$ as well as (iii). a node in the trace $\pmb{1}$ are subtracted as the last three graphs on the RHS of  \figref{Fig:Property-2}. Then only the cases where $b$ is a node in trace $\pmb{2}$ survive, thus the equality of \figref{Fig:Property-2} holds.  An explicit example for this transformation is given by \figref{Fig:example1}: According to the definition, the first graph on the RHS of \figref{Fig:Property-2} where two gravitons $l_1$ and $l_2$ belong to the set $L$ (the set of gravitons that are not in set $\{u_1, u_2\dots u_i, v_1, v_2, \dots v_j\}$) gives the equation of \figref{Fig:example1}, which can be verified by  moving the last four graphs on the RHS to the LHS.

 {\bf {Transformation-3:}}
\begin{figure}
\centering
\includegraphics[width=1.0\textwidth]{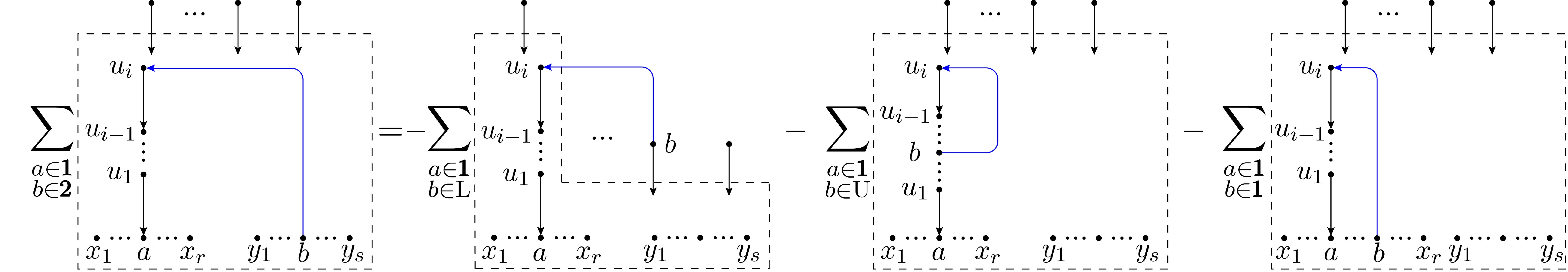}
\caption{Transformation-3 of spanning forests with the type-1 edge $e(b,u_i)$}
\label{Fig:Property-3}
\end{figure}
If the set $\{v_1,...,v_j\}$ is empty, the first term on the RHS of transformation-2 contains a factor
\bea
\Sl_{b}\Spbb{u_i,b}\frac{\Spaa{\zeta,u_i}\Spaa{\chi,b}}{\Spaa{\zeta,\chi}}=0,
\Label{Eq:Property-2.1}
\eea
where momentum conservation has been applied.  Transformation-2 then turns into the relation shown by \figref{Fig:Property-3}.

\subsection{The general proof} \label{Sec:GenProof}

To prove the formula (\ref{Eq:DoubleTraceMHV2}), we note that the RHS of (\ref{Eq:DoubleTraceMHV1}) has the general pattern of the LHS of \figref{Fig:Property-2} or \figref{Fig:Property-3}. One can therefore transform these factors according to transformation-2 and -3, whose terms on the RHS can be classified into the following three categories.

\begin{itemize}
\item {\bf Category-1}~~The first two graphs on the RHS of \figref{Fig:Property-2} and the first graph on the RHS of \figref{Fig:Property-3} have a common structure: all possible spanning forests while a chain structure ($v_j\to v_{j-1}\to...\to v_1$, $v_j\to v_{j-1}\to...\to b$ in the first two terms of \figref{Fig:Property-2} and the single node $b$ in \figref{Fig:Property-3}) is preserved are summed over and the starting node of this chain is connected to $u_i$ via a type-1 edge. All these graphs belong to category-1.

\item {\bf Category-2}~~In the third graph on the RHS of \figref{Fig:Property-2} and the second graph on the RHS of \figref{Fig:Property-3}, the chain $v_j\to v_{j-1}\to...\to v_1$ is planted at a node in the set of gravitons $\{u_1,...,u_i\}$ and  $v_j$ is connected to $u_i$ via a type-1 edge. Thus there is a loop structure with gravitons on it for a graph in this category.

\item {\bf Category-3}~~In the last graph on the RHS of \figref{Fig:Property-2} and the last graph on the RHS of \figref{Fig:Property-3}, all graphs where  the chain $v_j\to v_{j-1}\to...\to v_1$ pointing towards a gluon in trace $\pmb{1}$ are summed over. The starting node $v_j$ is also connected to $u_i$, via a type-1 edge. Thus there is a bridge between two gluons belonging to the trace $\pmb{1}$ in such a graph.
\end{itemize}
\begin{figure}
\centering
\includegraphics[width=0.77\textwidth]{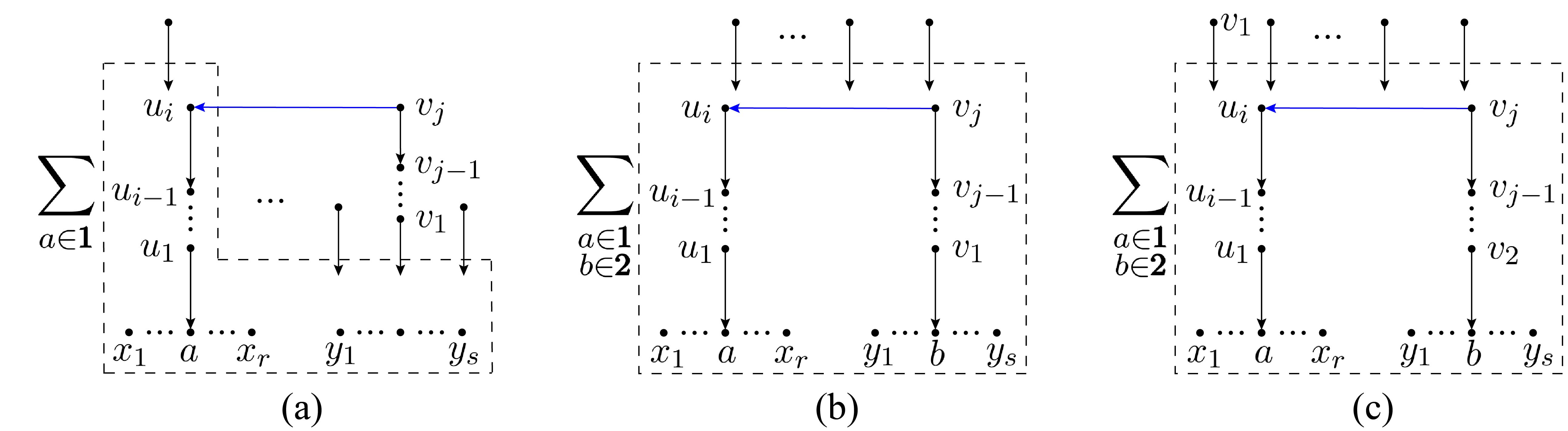}
\caption{The typical graph (a) in category-1 can only be obtained from spanning forests (b) and (c).}
\label{Fig:7}
\end{figure}
\begin{figure}
\centering
\includegraphics[width=0.74\textwidth]{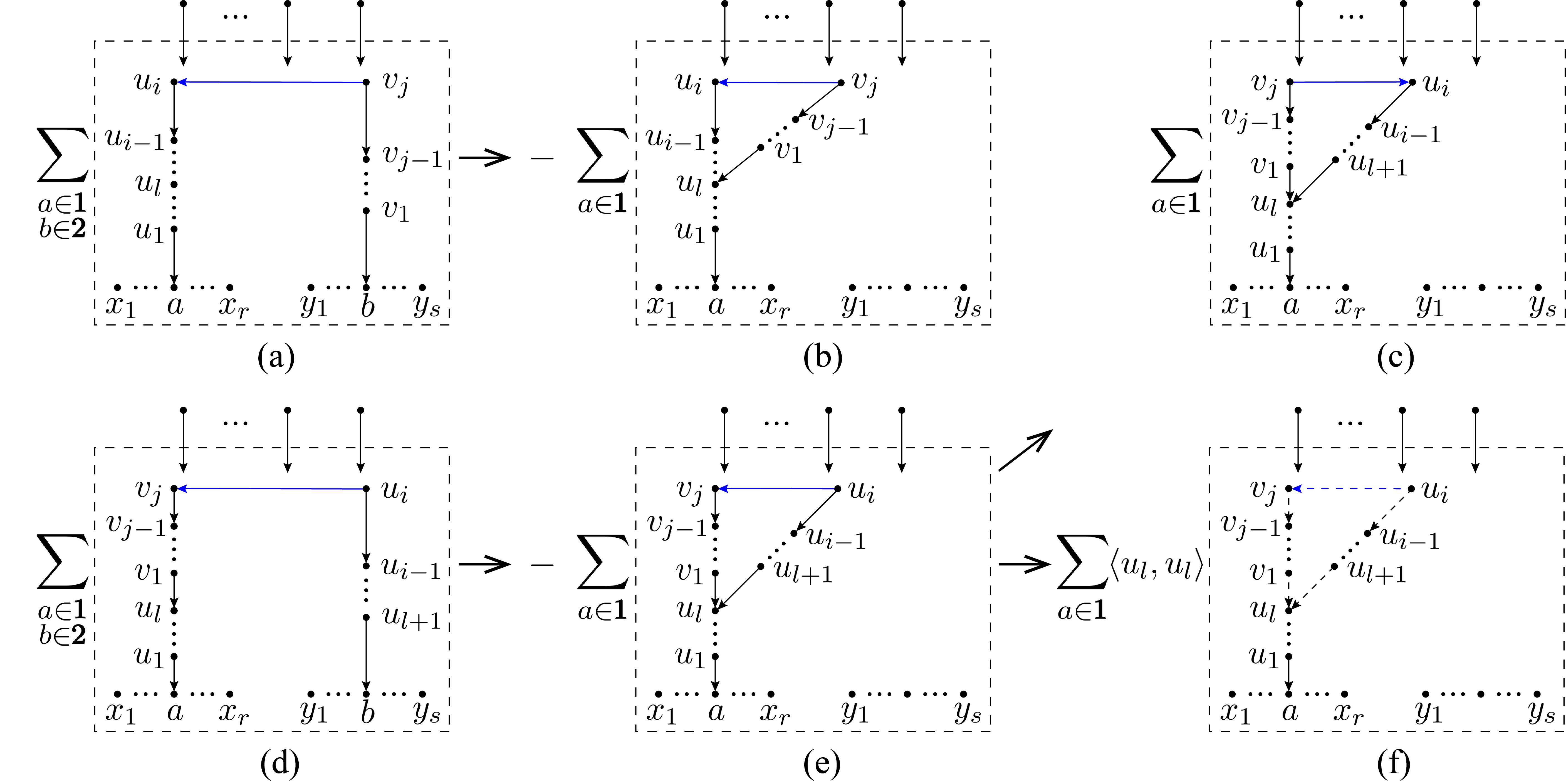}
\caption{A pair of graphs (b) and (e) in category-2, which come from the expansions of spanning forests (a) and (d), cancel each other out by applying transformation-1 to graph (e). Graphs (c) and (f) are the two graphs transformed from (e). }
\label{Fig:category2}
\end{figure}
\begin{figure}
\centering
\includegraphics[width=1.0\textwidth]{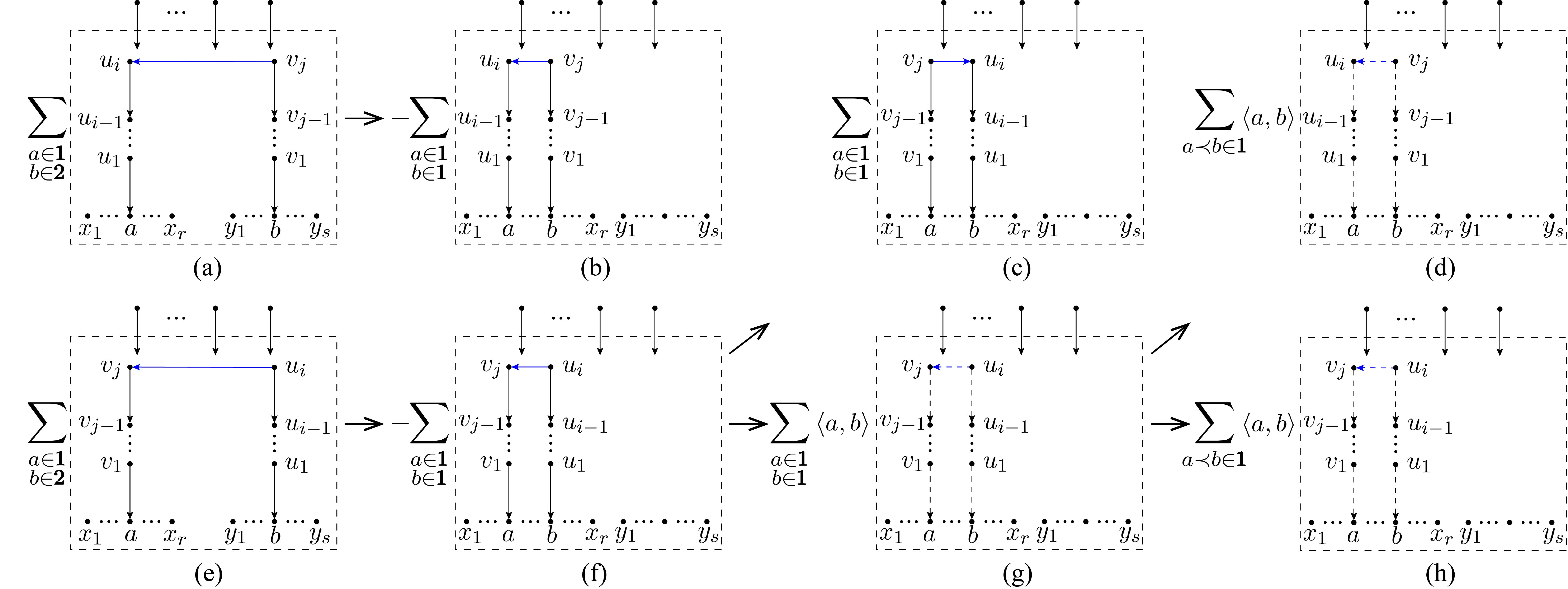}
\caption{A pair of graphs (b) and (f) in category-3, which are obtained from the expansions of spanning forests (a) and (e), only provide graphs (d) and (h). This is because after applying transformation-1 to graph (f), the resulting graph (c) cancels out graph (b), and the resulting graph (g) is further split into graphs (d) and (h).}
\label{Fig:category3}
\end{figure}
Now let us track back to the origins of these graphs in distinct categories and see the cancellations among graphs:

\begin{itemize}
\item  The typical graph of category-1 shown by \figref{Fig:7} (a) where the nodes on the bridge between $a$ and $b$ are $u_1,...,u_i,v_j,...,v_1$ in turn, can only be obtained from the expansions of \figref{Fig:7} (b) and (c). Since \figref{Fig:7} (a) is the first term in the expansion of \figref{Fig:7} (b) but the second term in the expansion of  \figref{Fig:7} (c)  when transformation-2 is applied, one such figure must appear twice with opposite signs and then cancel in pairs.

\item  Graphs of category-2 appear in pairs as shown by \figref{Fig:category2} (b) and (e), which respectively come from the decomposition of spanning forests (a) and (d). These two graphs are related to one another by exchanging   the roles between the chain structures $v_j\to v_{j-1}\to...\to v_1\to u_l$ and $u_i\to u_{i-1}\to...\to u_{l+1}\to u_l$. The graph (e) further splits into the two graphs (c) and (f), when transformation-1 is applied. The graph (c) cancels with (b), while the graph (f) has to vanish due to the antisymmetry of spinor products. Thus graphs of  category-2 have to cancel out.

\item  Graphs of category-3 also come in pairs as shown by \figref{Fig:category3} (b) and (f), which are obtained from the spanning forests (a) and (e). When applying transformation-1 to the graph (f), one gets the two graphs (c) and (g). The graph (c) cancels with (b), while the graph (g) is further expanded into graphs (d) and (h).  This is because the summation over $a\in \pmb{1}, b\in \pmb{1}$ can split as
\bea
 \Sl_{\substack{a\in\pmb{1}\\ b\in \pmb{1}}}\to  \Sl_{\substack{a\prec b\\ a,b\in \pmb{1}}}+\Sl_{\substack{b\prec a\\ a,b\in \pmb{1}}}.
\eea
When  $a$, $b$ are renamed by $b$, $a$, respectively, the second summation on the RHS turns into a summation over $a\prec b~(a,b\in \pmb{1})$ but the structure between $a$ and $b$ is reflected, as shown by the graph (d) in \figref{Fig:category3} (the direction of the arrow line from $u_i$ to $v_j$ is adjusted by absorbing a minus into the coefficient $\Spaa{b,a}=-\Spaa{a,b}$).
\end{itemize}
\begin{figure}
\centering
\includegraphics[width=1.0\textwidth]{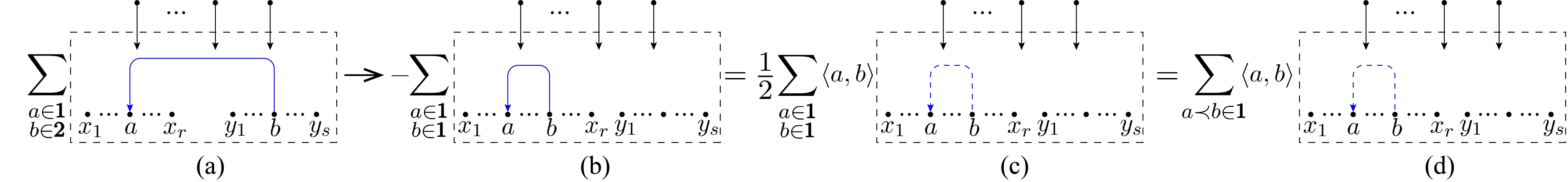}
\caption{The transformation of the boundary case that gluons $a$ and $b$ are directly connected via a type-1 edge}
\label{Fig:boundary}
\end{figure}

 There is a boundary case of category-3 when both sets $\{u_1,...,u_i\}$ and $\{v_1,...,v_j\}$ are empty, as shown by the graph (b) in \figref{Fig:boundary}, which is obtained after applying transformation-3 to spanning forest (a). When we apply transformation-1 to this graph (b), the LHS of \figref{Fig:Property-1} is same with the second term on the RHS (upto a minus), thus the LHS is half of the first term on the RHS, as shown by the first equality of  \figref{Fig:boundary}. The summation over all $a, b\in \pmb{1}$ is further written as twice of the summation over $a,b$ ($a, b\in \pmb{1}, a\prec b$).

To sum up, only the graphs of structures \figref{Fig:category3} (d), (h) and \figref{Fig:boundary} (d) survive after cancellations (An explicit one-graviton example is given in \appref{app:one-graviton}.). These graphs are those characterizing \eqref{Eq:DoubleTraceMHV2}, hence proof of the equivalence between \eqref{Eq:DoubleTraceMHV1} and \eqref{Eq:DoubleTraceMHV2} is completed.

\section{CHY approach to double-trace MHV amplitudes with two negative-helicity gluons}\label{SE:CHYApproach}

In this section, we evaluate the double-trace MHV amplitude in four dimensions by CHY formula directly. We  straightforwardly substitute the MHV solution into the (integrated) CHY formula for double-trace amplitudes, then show that this sector reduces into the formula (\ref{Eq:DoubleTraceMHV2}).

\subsection{A review of the CHY formula for multi-trace EYM amplitudes}

In the Cachazo-He-Yuan (CHY) framework, an $n$-point scattering amplitude $M_n$  can be given by the following (integrated) formula
\bea
M_n=\Sl_{\{\sigma\}\in sol}\frac{\mathcal{I}_n}{\text{det}\,'[\Phi]}.\Label{Eq:IntegratedCHY}
\eea
In the above expression:
\begin{itemize}
\item One have summed over solutions $\{\sigma_a\}$ ($a=1,...,n$) of the following scattering equations for scattering variables $\{z_a\}$
\bea
 \Sl^n_{\substack{b=1\\b\ne a}}\frac{s_{ab}}{z_{ab}}=0 \,,\,\,a\in\{1,2,\dots,n\}\, ,
\eea
where  $s_{ab}=2k_a\cdot k_b$ and $z_{ab}=z_a-z_b$. In four dimensions, the scattering equations have a special solution (which is called MHV solution)
\bea
{\sigma^{\text{MHV}}_{a}}=\frac{\Spaa{a,\eta}\Spaa{\theta,\zeta}}{\Spaa{a,\zeta}\Spaa{\theta,\eta}}\,,\Label{Eq:MHVSolution}
\eea
where the spinors $\eta$, $\theta$ and $\zeta$ represent the M$\ddot{\text{o}}$bius freedom of scattering equations.

\item The reduced determinant $\text{det}\,'[\Phi]$ is defined by
\bea
\text{det}\,'[\Phi]\equiv perm(ijk)perm(pqr)\frac{\text{det}[\Phi^{ijk}_{pqr}]}{\sigma_{ij}\sigma_{jk}\sigma_{ki}\sigma_{pq}\sigma_{qr}\sigma_{rp}},
\eea
where $perm(ijk)$ is the signature of the permutation that moves the standard ordering $(1,2,...,n)$ to $(i,j,k,...)$ and keeps $(...)$ ascending. The $n\times n$ matrix $\Phi$ is given by
\bea
\Phi_{ab}=\Bigg\{
            \begin{array}{cc}
              \frac{s_{ab}}{\sigma^2_{ab}} &, a\ne b \\
               -\Sl_{c\ne a}\frac{s_{ac}}{\sigma^2_{ac}} &,a=b \\
            \end{array}\,,
\eea
and $\Phi^{ijk}_{pqr}$ is the submatrix obtained by removing the $(i,j,k)$-th row and the $(p,q,r)$-th column from $\Phi$.

\item The CHY integrand ${\cal I}_n$ relies on theories. In EYM theory, the CHY integrand $\mathcal{I}_{\text{EYM}}$ for  color-odered $m$-trace amplitudes have the following form
\bea
\mathcal{I}_{\text{EYM}}(\pmb{1}|\dots |\pmb{m}\Vert\mathsf{H})=\underbrace{\big(\mathcal{C}_{\pmb{1}}\dots\mathcal{C}_{\pmb{m}}\,{\Sl}'_{\{a,b\}}P_{\{a,b\}}\big)}_{\mathcal{I}_L}\underbrace{\text{Pf}\,'[\Psi]}_{\mathcal{I}_R}\, ,\Label{Eq:EYMIntegrand}
\eea
where $\mathcal{C}_{\pmb{i}}\equiv\frac{1}{z_{a_1a_2}z_{a_2a_3}\dots z_{a_sa_1}}$ is the Parke-Taylor factor  for the gluon trace $\pmb{i}=\{a_1,a_2,...,a_s\}$ and the reduced Pfaffian $\text{Pf}\,'[\Psi]$ is defined as
\bea
\text{Pf}\,'[\Psi]\equiv\frac{perm(ij)}{z_{ij}}\text{Pf}[\Psi_{ij}^{ij}],\,(1\le i<j\le n).
\eea
The $\Psi$ is an $2n\times 2n$ antisymmetric matrix with the following structure
\[\Psi=\begin{pmatrix}
A&-C^T\\
C&B\\
\end{pmatrix},
\]
where the three blocks $A$, $B$ and $C$ are $n\times n$ matrices defined as follows
\bea
A_{ab}=\Bigg\{
            \begin{array}{cc}
              \frac{k_a\cdot k_b}{z_{ab}} &, a\ne b \\
               0 &,a=b \\
            \end{array}\,,~~
B_{ab}=\Bigg\{
            \begin{array}{cc}
               \frac{\epsilon_a\cdot \epsilon_b}{z_{ab}} &, a\ne b \\
               0 &,a=b \\
            \end{array}\,,~~
C_{ab}=\Bigg\{
            \begin{array}{cc}
              \frac{\epsilon_a\cdot k_b}{z_{ab}} &, a\ne b \\
              -\Sl_{c\ne a}\frac{\epsilon_a\cdot k_c}{z_{ac}} &,a=b \\
            \end{array}\,.
\Label{Eq:3}
\eea
The $\epsilon_a$ denotes half polarization of the graviton $a$. In the integrand (\ref{Eq:EYMIntegrand}),
\bea
&{\Sl}'_{\{a,b\}}P_{\{a,b\}}=\Sl_{\substack{a_1\prec b_1 \in \pmb{1}\\\dots\\a_{m-1}\prec b_{m-1} \in \pmb{m-1}}}sgn(\{a,b\})z_{a_1b_1}\dots z_{a_{m-1}b_{m-1}}\text{Pf}\,[\Psi]_{\mathsf{H},a_1,b_1,\dots,a_{m-1},b_{m-1};\mathsf{H}}\,,\Label{Eq:EYMIntegrand1}
\eea
 where $\{a,b\}$ represents $\{a_1,b_1,\dots,a_{m-1},b_{m-1}\}$.
\end{itemize}
It is worth pointing out that Pfaffians can be defined recursively
\bea
\text{Pf}\,[\Psi]=\Sl^n_{j=1,j\ne i}(-1)^{i+j+1+\theta(i-j)}\,x_{ij} \,\text{Pf}\,[\Psi^{ij}_{ij}]\,\, .
\Label{Eq:Pfaffianexpansion2}
\eea
A simple but helpful property is the following: if a  $2t\times 2t$ antisymmetric matrix has the  form
\bea
\Psi=
\begin{bmatrix}
A      & -C^T      \\
C      &  O
\end{bmatrix}\,,\Label{Eq:Pfaffianexpansion0}
\eea
where both $A$ and $C$ are $t\times t$ blocks, $O$ is an all-zero block. The Pfaffian of (\ref{Eq:Pfaffianexpansion0}) can be reduced as
\bea
\text{Pf}\,[\Psi]=(-1)^{\frac{t(t+1)}{2}}\text{det}\,[C].
\Label{Eq:Pfaffianexpansion1}
\eea

\subsection{Evaluating the double-trace $(g^-,g^-)$-amplitudes by CHY formula}
Now let us evaluate the double-trace amplitude with  $(g^-,g^-)$-configuration by CHY formula straightforwardly. When the EYM integrand (\ref{Eq:EYMIntegrand}) for double-trace amplitude is substituted into the integrated expression (\ref{Eq:IntegratedCHY}), the amplitude becomes\footnote{The sign in (\ref{Eq:EYMIntegrand1}) is absorbed if the $a$, $b$ in the Pfaffian are arranged in canonical order.}
\bea
A(x_1,\ldots, x_r|y_1,\ldots,y_s\Vert \mathsf{H})=\Sl_{\{\sigma\}\in sol}\frac{1}{\sigma_{x_1x_2}\dots\sigma_{x_rx_1}}\frac{1}{\sigma_{y_1y_2}\dots\sigma_{y_sy_1}}\Sl_{a\prec b\in\pmb{1}}\,\sigma_{ab}\,\text{Pf}\,[\Psi]_{\mathsf{H},a,b;\mathsf{H}}\,\frac{\text{Pf}\,'[\Psi]}{\text{det}\,'[\Phi]}.
\eea
For the sector supported by the MHV solution (\ref{Eq:MHVSolution}), we have
\bea
A^{\text{MHV}}(x_1,\ldots, x_r|y_1,\ldots,y_s\Vert \mathsf{H})&=&\frac{(-1)^{\frac{n^2-3n+8}{2}}(\sqrt{2})^n\Spaa{i,j}^4}{F^nP^2_\zeta}\nn
&&~~~~~~~~\times\frac{1}{\sigma_{x_1x_2}\dots\sigma_{x_rx_1}}\frac{1}{\sigma_{y_1y_2}\dots\sigma_{y_sy_1}}\Sl_{a\prec b\in\pmb{1}}\,\sigma_{ab}\,\text{Pf}\,[\Psi]_{\mathsf{H},a,b;\mathsf{H}},
\eea
where we omit the superscript of $\sigma^{\text{MHV}}_a$ for convenience. In the above expression, we have used the property introduced in \cite{{Du:2016wkt}}
\bea
\frac{\text{Pf}\,'[\Psi]}{\text{det}\,'[\Phi]}=\frac{(-1)^{\frac{n^2-3n+8}{2}}(\sqrt{2})^n\Spaa{i,j}^4}{F^nP^2_\zeta}\,,
\eea
where the $i$, $j$ are the two negative-helicity particles, and
\bea
F=\frac{\Spaa{\theta,\eta}}{\Spaa{\theta,\zeta}\Spaa{\eta,\zeta}}\,,P_\zeta=\prod_{a=1}^n\Spaa{a,\zeta}\,.
\eea
The submatrix $[\Psi]_{\mathsf{H},a,b;\mathsf{H}}$ of $\Psi$ consists of three blocks $A'$, $B'$ and $C'$. The indices of square matrices $A'$ and $B'$ correspondingly take values as  (i). elements of $\mathsf{H}\cup\{a,b\}$ in order, and (ii). elements in $\mathsf{H}$. The block $C'$ is the submatrix of $C$ with row and column indices taking values in  $\mathsf{H}$ and $\mathsf{H}\cup\{a,b\}$ respectively.

In the case that the negative-helicity particles $i$, $j$ are gluons, the Lorentz contraction of momenta and polarizations in $\text{Pf}\,[\Psi]_{\mathsf{H},a,b;\mathsf{H}}$ are expressed by
\bea
k_a\cdot k_b=\Spaa{a,b}\Spbb{b,a}\,,k_a\cdot\epsilon_b^+(\xi)=\frac{\Spaa{\xi,a}\Spbb{a,b}}{\Spaa{\xi,b}}\,,k_a\cdot\epsilon_b^-(q)=\frac{\Spaa{a,b}\Spbb{q,a}}{\Spbb{q,b}}\,,\epsilon_a^+(\xi)\cdot\epsilon_b^+(\xi)=0\,
\eea
where we choose the reference spinor for all positive-helicity particles as $\xi$, while that for the negative-helicity particles as $q$.
When $(\Spaa{h_1,\zeta}\dots\Spaa{h_t,\zeta})^2\Spaa{a,\zeta}\Spaa{b,\zeta}F^{2t+2}$  and $(\Spaa{h_1,\zeta}\dots\Spaa{h_t,\zeta})^2\Spaa{a,\zeta}\Spaa{b,\zeta}$ are extracted out from rows and columns in $\text{Pf}\,[\Psi]_{\mathsf{H},a,b;\mathsf{H}}$, respectively, the amplitude turns into
\bea
A^{(g^-_i,g^-_j)}(x_1,\ldots, x_r|y_1,\ldots,y_s\Vert \mathsf{H})\propto \frac{\Spaa{g_i,g_j}^4}{\left(x_1,\dots,x_r\right)\left(y_1,\dots,y_s\right)}\left(\Sl_{a\prec b\in\pmb{1}}\,\Spaa{a,b} \text{Pf}\,[\W{\Psi}]_{\mathsf{H},a,b;\mathsf{H}}\right)\,,
\Label{Eq:DoubleTraceMHV3}
\eea
where
\bea
\W{\Psi}=\left(\begin{array}{cc}
\W{A}&-\W{C}^T\\
\W{C}&\W{B}\\
\end{array}\right)
,~~\,\W{A}_{ab}=\Bigg\{
            \begin{array}{cc}
              \Spbb{b,a} &, a\ne b \\
               0 &,a=b \\
            \end{array}\,,~~\,
\W{B}_{ab}=0,~~\,
\W{C}_{ab}=\Bigg\{
            \begin{array}{cc}
              -\frac{\Spaa{\xi,b}\Spbb{b,a}}{\Spaa{\xi,a}\Spaa{b,a}} &, a\ne b \\
               \Sl^n_{c=1,c\ne a}\frac{\Spaa{\xi,c}\Spaa{\zeta,c}\Spbb{c,a}}{\Spaa{\xi,a}\Spaa{\zeta,a}\Spaa{c,a}} &,a=b \\
            \end{array}\,.\Label{Eq:ABCtilde}
\eea
In the coming subsection, we prove that the summation in \eqref{Eq:DoubleTraceMHV3} can be expanded into the expression inside the square brackets of \eqref{Eq:DoubleTraceMHV2}. Then the result from the MHV sector (i.e. the sector supported by the MHV solution) of CHY formula precisely reduces into the expansion formula (\ref{Eq:DoubleTraceMHV2}) and is therefore equivalent to the spanning forest form (\ref{Eq:DoubleTraceMHV1}).

\subsection{The expansion of $\text{Pf}\,[\W{\Psi}]_{\mathsf{H},a,b;\mathsf{H}}$}

In this subsection, we expand the $\text{Pf}\,[\W{\Psi}]_{\mathsf{H},a,b;\mathsf{H}}$ and show that this expansion results in the expression in the  square brackets of \eqref{Eq:DoubleTraceMHV2}. We first provide the examples with one and two gravitons, then the general proof.

\subsubsection*{One-graviton case}
\begin{figure}
\centering
\includegraphics[width=0.85\textwidth]{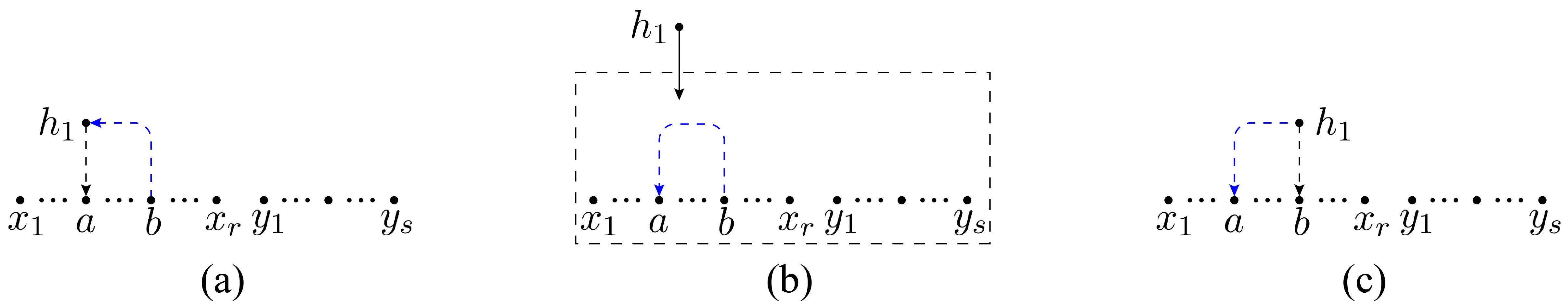}
\caption{Three graphs corresponding to the three terms of the expansion of one-graviton Pfaffian $\text{Pf}\,[\W{\Psi}]_{h_1,a,b;h_1}$}
\label{Fig:14}
\end{figure}
The simplest example is the Pfaffian  $\text{Pf}\,[\W{\Psi}]_{h_1,a,b;h_1}$ involving one graviton $h_1$, which is explicitly written as
\bea
\text{Pf}\,
\begin{bmatrix}
0 & \Spbb{a,h_1} & \Spbb{b,h_1} & -\phi_{h_1}      \\
\Spbb{h_1,a} & 0 & \Spbb{b,a} & \frac{\Spaa{\xi,a}\Spbb{a,h_1}}{\Spaa{\xi,h_1}\Spaa{a,h_1}}     \\
\Spbb{h_1,b} & \Spbb{a,b} & 0 & \frac{\Spaa{\xi,b}\Spbb{b,h_1}}{\Spaa{\xi,h_1}\Spaa{b,h_1}}     \\
\phi_{h_1} & -\frac{\Spaa{\xi,a}\Spbb{a,h_1}}{\Spaa{\xi,h_1}\Spaa{a,h_1}}  & -\frac{\Spaa{\xi,b}\Spbb{b,h_1}}{\Spaa{\xi,h_1}\Spaa{b,h_1}} & 0
\end{bmatrix}.
\eea
According to the recursive expression  (\ref{Eq:Pfaffianexpansion2}), the Pfaffian can be given by
\bea
\text{Pf}\,[\W{\Psi}]_{h_1,a,b;h_1}=(-1)\Big(\Spbb{b,h_1}\frac{\Spaa{\xi,a}\Spbb{a,h_1}}{\Spaa{\xi,h_1}\Spaa{a,h_1}}+\Spbb{b,a}\phi_{h_1} +\Spbb{h_1,a}\frac{\Spaa{\xi,b}\Spbb{b,h_1}}{\Spaa{\xi,h_1}\Spaa{b,h_1}}\Big)\,,
\eea
where $\phi_{h_1}=\Sl^n_{c=1,c\ne h_1}\frac{\Spaa{\xi,c}\Spaa{\zeta,c}\Spbb{c,h_1}}{\Spaa{\xi,h_1}\Spaa{\zeta,h_1}\Spaa{c,h_1}}$. The above three terms correspond to the three graphs in \figref{Fig:14} which describe the expansion formula (\ref{Eq:DoubleTraceMHV2}) with one graviton.

\subsubsection*{Two-graviton case}

The $\text{Pf}\,[\W{\Psi}]_{h_1,h_2,a,b;h_1,h_2}$ with two gravitons $h_1,h_2$ are explicitly given by
\bea
&\text{Pf}&\,
\begin{bmatrix}
0 & \Spbb{h_2,h_1} & \Spbb{a,h_1} & \Spbb{b,h_1} & -\phi_{h_1} & \frac{\Spaa{\xi,h_1}\Spbb{h_1,h_2}}{\Spaa{\xi,h_2}\Spaa{h_1,h_2}}    \\
\Spbb{h_1,h_2} & 0 & \Spbb{a,h_2} & \Spbb{b,h_2} & \frac{\Spaa{\xi,h_2}\Spbb{h_2,h_1}}{\Spaa{\xi,h_1}\Spaa{h_2,h_1}} & -\phi_{h_2}    \\
\Spbb{h_1,a} & \Spbb{h_2,a} & 0 & \Spbb{b,a} & \frac{\Spaa{\xi,a}\Spbb{a,h_1}}{\Spaa{\xi,h_1}\Spaa{a,h_1}} & \frac{\Spaa{\xi,a}\Spbb{a,h_2}}{\Spaa{\xi,h_2}\Spaa{a,h_2}}    \\
\Spbb{h_1,b} & \Spbb{h_2,b} & \Spbb{a,b} & 0 & \frac{\Spaa{\xi,b}\Spbb{b,h_1}}{\Spaa{\xi,h_1}\Spaa{b,h_1}} & \frac{\Spaa{\xi,b}\Spbb{b,h_2}}{\Spaa{\xi,h_2}\Spaa{b,h_2}}    \\
\phi_{h_1} & -\frac{\Spaa{\xi,h_2}\Spbb{h_2,h_1}}{\Spaa{\xi,h_1}\Spaa{h_2,h_1}} &  -\frac{\Spaa{\xi,a}\Spbb{a,h_1}}{\Spaa{\xi,h_1}\Spaa{a,h_1}}  & -\frac{\Spaa{\xi,b}\Spbb{b,h_1}}{\Spaa{\xi,h_1}\Spaa{b,h_1}} & 0 & 0 \\
-\frac{\Spaa{\xi,h_1}\Spbb{h_1,h_2}}{\Spaa{\xi,h_2}\Spaa{h_1,h_2}} & \phi_{h_2}  &  -\frac{\Spaa{\xi,a}\Spbb{a,h_2}}{\Spaa{\xi,h_2}\Spaa{a,h_2}}  & -\frac{\Spaa{\xi,b}\Spbb{b,h_2}}{\Spaa{\xi,h_2}\Spaa{b,h_2}} & 0 & 0
\end{bmatrix}\nn
&=&\Spbb{b,h_1}T_1+\Spbb{b,h_2}T_2-\Spbb{b,a}T_3
+\frac{\Spaa{\xi,b}\Spbb{b,h_1}}{\Spaa{\xi,h_1}\Spaa{b,h_1}}T_4+\frac{\Spaa{\xi,b}\Spbb{b,h_2}}{\Spaa{\xi,h_2}\Spaa{b,h_2}}T_5,
\Label{Eq:Pfaffian2}
\eea
where the recursive expansion of Pfaffian (\ref{Eq:Pfaffianexpansion2}) is applied again. The $T_1$-$T_5$ are reduced as follows:
\begin{figure}
\centering
\includegraphics[width=0.85\textwidth]{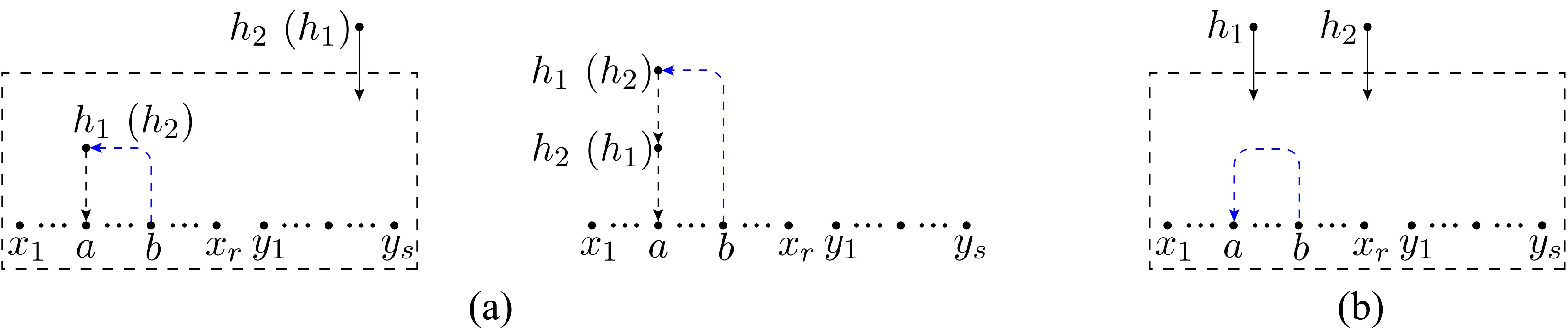}
\caption{(a): All two-graviton graphs with  $h_1$ ($h_2$) as starting node pointing towards gluon $a$ via one or two type-4 edges, (b): The spanning forests for the two-graviton Hodges determinant, accompanied by a type-3 edge $e(b,a)$}
\label{Fig:15a}
\end{figure}
\begin{itemize}
\item $T_1$ and $T_2$~~When the property (\ref{Eq:Pfaffianexpansion1}) is applied, the factors $T_1$ and $T_2$ in the first two terms on the RHS of (\ref{Eq:Pfaffian2}) are given by two determinants
\bea
T_1=\begin{vmatrix}
\phi_{h_2} &  -\frac{\Spaa{\xi,a}\Spbb{a,h_2}}{\Spaa{\xi,h_2}\Spaa{a,h_2}}  \\
-\frac{\Spaa{\xi,h_2}\Spbb{h_2,h_1}}{\Spaa{\xi,h_1}\Spaa{h_2,h_1}} & -\frac{\Spaa{\xi,a}\Spbb{a,h_1}}{\Spaa{\xi,h_1}\Spaa{a,h_1}}
\end{vmatrix},\,~~~~~
T_2=\begin{vmatrix}
\phi_{h_1} &  -\frac{\Spaa{\xi,a}\Spbb{a,h_1}}{\Spaa{\xi,h_1}\Spaa{a,h_1}}  \\
-\frac{\Spaa{\xi,h_1}\Spbb{h_1,h_2}}{\Spaa{\xi,h_2}\Spaa{h_1,h_2}} & -\frac{\Spaa{\xi,a}\Spbb{a,h_2}}{\Spaa{\xi,h_2}\Spaa{a,h_2}}
\end{vmatrix}.
\eea
The expansion of the determinant $T_1$ (and $T_2$), together with the prefactor $\Spbb{b,h_1}$ ($\Spbb{b,h_2}$), reproduce the graphs with type-3 edge $e(b,h_1)$ ($e(b,h_2)$) in \figref{Fig:15a} (a).
\item $T_3$~~The factor $T_3$ in the third term on the RHS of (\ref{Eq:Pfaffian2}) has the following form
\bea
T_3=\begin{vmatrix}
\phi_{h_1} &  -\frac{\Spaa{\xi,h_2}\Spbb{h_2,h_1}}{\Spaa{\xi,h_1}\Spaa{h_2,h_1}} \\
-\frac{\Spaa{\xi,h_1}\Spbb{h_1,h_2}}{\Spaa{\xi,h_2}\Spaa{h_1,h_2}} & \phi_{h_2}
\end{vmatrix}.
\eea
Since the $n\times n$ Hodges determinant \cite{Hodges:2012ym} is defined as
\bea
\phi_{ab}=\frac{\Spbb{ab}}{\Spaa{ab}}\,\, (a\neq b)\,\,,~~~~~~ \phi_{aa}=-\Sl^n_{c=1,c\ne a}\frac{\Spaa{\xi,c}\Spaa{\zeta,c}\Spbb{c,a}}{\Spaa{\xi,a}\Spaa{\zeta,a}\Spaa{c,a}}\,\,(a=b )\,\, ,\Label{Eq:Hodgesdeterminant}
\eea
where the reference spinors $\xi$ and $\zeta$ are arbitrarily chosen, the determinant $T_3$ is equivalent to the Hodges determinant with two gravitons $h_1$ and $h_2$. The spanning forest \cite{Nguyen:2009jk,Feng:2012sy} corresponding to the Hodges determinant, together with the factor $-\Spbb{b,a}$, is shown by \figref{Fig:15a} (b).
\begin{figure}
\centering
\includegraphics[width=0.85\textwidth]{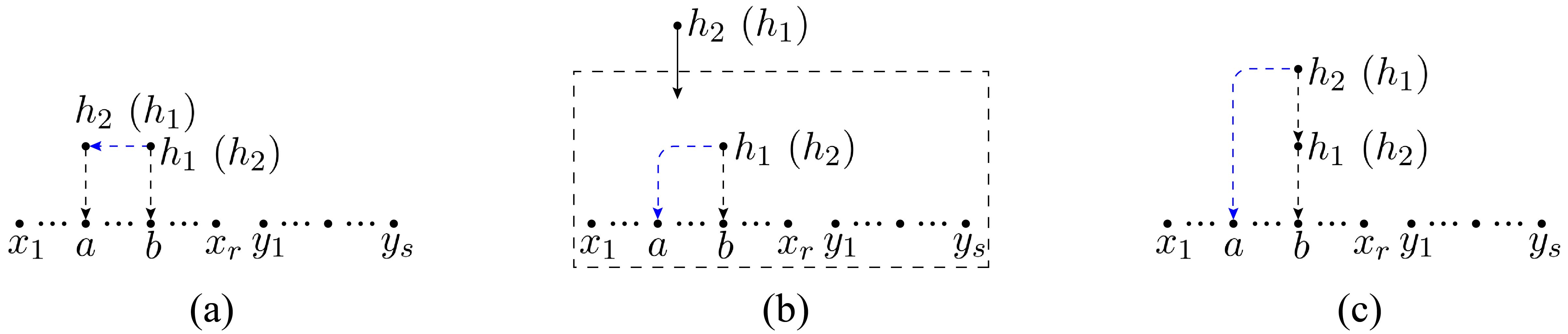}
\caption{All two-graviton graphs with $h_1$ ($h_2$) directly connected to $b$ via a type-4 edge}
\label{Fig:15b}
\end{figure}
\item $T_4$ and $T_5$~~The last two factors $T_4$ and $T_5$ on the RHS of (\ref{Eq:Pfaffian2}) are given by
\bea
T_4=\text{Pf}\,
\begin{bmatrix}
0 & \Spbb{a,h_2} & \Spbb{h_1,h_2} & -\phi_{h_2}      \\
\Spbb{h_2,a} & 0 & \Spbb{h_1,a} & \frac{\Spaa{\xi,a}\Spbb{a,h_2}}{\Spaa{\xi,h_2}\Spaa{a,h_2}}     \\
\Spbb{h_2,h_1} & \Spbb{a,h_1} & 0 & \frac{\Spaa{\xi,h_1}\Spbb{h_1,h_2}}{\Spaa{\xi,h_2}\Spaa{h_1,h_2}}     \\
\phi_{h_2} & -\frac{\Spaa{\xi,a}\Spbb{a,h_2}}{\Spaa{\xi,h_2}\Spaa{a,h_2}}  & -\frac{\Spaa{\xi,h_1}\Spbb{h_1,h_2}}{\Spaa{\xi,h_2}\Spaa{h_1,h_2}} & 0
\end{bmatrix},~~~T_5=T_4(h_1\leftrightarrow h_2).
\eea
Note that $T_4$ (and $T_5$) can be obtained by performing the replacements $h_1\to h_2$, $b\to h_1$ ($b\to h_2$ ) simultaneously on the Pfaffian $\text{Pf}\,[\W{\Psi}]_{h_1,a,b;h_1}$ with one graviton, which was already expanded as \figref{Fig:14}. When multiplied by the corresponding factor in \eqref{Eq:Pfaffian2} (i.e. the edge between $h_1$ ($h_2$) and $b$),  $T_4$ (and $T_5$) becomes terms characterized by \figref{Fig:15b}.

\end{itemize}
To summarize, all terms in the expansion of $\text{Pf}\,[\W{\Psi}]_{h_1,h_2,a,b;h_1,h_2}$ correspond to the graphs \figref{Fig:15a} and \figref{Fig:15b}, which are all contributions of (\ref{Eq:DoubleTraceMHV2}) for amplitude with two gravitons.

\subsubsection*{General expansion}
The $t$-graviton Pfaffian $\text{Pf}\,[\W{\Psi}]_{\mathsf{H},a,b;\mathsf{H}}$ for double-trace amplitudes is explicitly displayed as
\bea
&\text{Pf}&\,
\footnotesize{
\left[\begin{array}{cccccc|cccc}
0 & \Spbb{h_2,h_1} & \cdots & \Spbb{h_t,h_1} & \Spbb{a,h_1} & \Spbb{b,h_1} & -\phi_{h_1} & \frac{\Spaa{\xi,h_1}\Spbb{h_1,h_2}}{\Spaa{\xi,h_2}\Spaa{h_1,h_2}} & \cdots & \frac{\Spaa{\xi,h_1}\Spbb{h_1,h_t}}{\Spaa{\xi,h_t}\Spaa{h_1,h_t}}      \\
\Spbb{h_1,h_2} & 0 & \cdots & \Spbb{h_t,h_2} & \Spbb{a,h_2} & \Spbb{b,h_2} & \frac{\Spaa{\xi,h_2}\Spbb{h_2,h_1}}{\Spaa{\xi,h_1}\Spaa{h_2,h_1}} & -\phi_{h_2} & \cdots & \frac{\Spaa{\xi,h_2}\Spbb{h_2,h_t}}{\Spaa{\xi,h_t}\Spaa{h_2,h_t}}   \\
\vdots & \vdots & \ddots & \vdots & \vdots & \vdots & \vdots & \vdots & \ddots & \vdots   \\
\Spbb{h_1,h_t} & \Spbb{h_1,h_t} & \cdots & 0 & \Spbb{a,h_t} & \Spbb{b,h_t} & \frac{\Spaa{\xi,h_t}\Spbb{h_t,h_1}}{\Spaa{\xi,h_1}\Spaa{h_t,h_1}} & \frac{\Spaa{\xi,h_t}\Spbb{h_t,h_2}}{\Spaa{\xi,h_2}\Spaa{h_t,h_2}} & \cdots & -\phi_{h_t}   \\
\Spbb{h_1,a} & \Spbb{h_2,a} & \cdots & \Spbb{h_t,a} & 0 & \Spbb{b,a} & \frac{\Spaa{\xi,a}\Spbb{a,h_1}}{\Spaa{\xi,h_1}\Spaa{a,h_1}} & \frac{\Spaa{\xi,a}\Spbb{a,h_2}}{\Spaa{\xi,h_2}\Spaa{a,h_2}} & \cdots & \frac{\Spaa{\xi,a}\Spbb{a,h_t}}{\Spaa{\xi,h_t}\Spaa{a,h_t}}   \\
\Spbb{h_1,b} & \Spbb{h_2,b} & \cdots & \Spbb{h_t,b} & \Spbb{a,b} & 0 & \frac{\Spaa{\xi,b}\Spbb{b,h_1}}{\Spaa{\xi,h_1}\Spaa{b,h_1}} & \frac{\Spaa{\xi,b}\Spbb{b,h_2}}{\Spaa{\xi,h_2}\Spaa{b,h_2}} & \cdots & \frac{\Spaa{\xi,b}\Spbb{b,h_t}}{\Spaa{\xi,h_t}\Spaa{b,h_t}}   \\ \hline
\phi_{h_1} & -\frac{\Spaa{\xi,h_2}\Spbb{h_2,h_1}}{\Spaa{\xi,h_1}\Spaa{h_2,h_1}} & \cdots & \frac{\Spaa{\xi,h_t}\Spbb{h_t,h_1}}{\Spaa{\xi,h_1}\Spaa{h_t,h_1}} &  -\frac{\Spaa{\xi,a}\Spbb{a,h_1}}{\Spaa{\xi,h_1}\Spaa{a,h_1}}  & -\frac{\Spaa{\xi,b}\Spbb{b,h_1}}{\Spaa{\xi,h_1}\Spaa{b,h_1}} & 0 & 0 & \cdots & 0   \\
 -\frac{\Spaa{\xi,h_1}\Spbb{h_1,h_2}}{\Spaa{\xi,h_2}\Spaa{h_1,h_2}} & \phi_{h_2} &\cdots & \frac{\Spaa{\xi,h_t}\Spbb{h_t,h_2}}{\Spaa{\xi,h_2}\Spaa{h_t,h_2}} &  -\frac{\Spaa{\xi,a}\Spbb{a,h_2}}{\Spaa{\xi,h_2}\Spaa{a,h_2}}  & -\frac{\Spaa{\xi,b}\Spbb{b,h_2}}{\Spaa{\xi,h_2}\Spaa{b,h_2}} & 0 & 0 & \cdots & 0   \\
\vdots & \vdots & \ddots & \vdots & \vdots & \vdots & \vdots & \vdots & \ddots & \vdots   \\
-\frac{\Spaa{\xi,h_1}\Spbb{h_1,h_t}}{\Spaa{\xi,h_t}\Spaa{h_1,h_t}} & -\frac{\Spaa{\xi,h_2}\Spbb{h_2,h_t}}{\Spaa{\xi,h_t}\Spaa{h_2,h_t}} & \cdots & \phi_{h_t}  &  -\frac{\Spaa{\xi,a}\Spbb{a,h_t}}{\Spaa{\xi,h_t}\Spaa{a,h_t}}  & -\frac{\Spaa{\xi,b}\Spbb{b,h_t}}{\Spaa{\xi,h_t}\Spaa{b,h_t}} & 0 & 0 & \cdots & 0
\end{array}\right]}.\nn
\eea
Inspired by the two-graviton example, we expand this Pfaffian  by entries on the $(t+2)$-th row into three parts:
\begin{figure}
\centering
\includegraphics[width=0.8\textwidth]{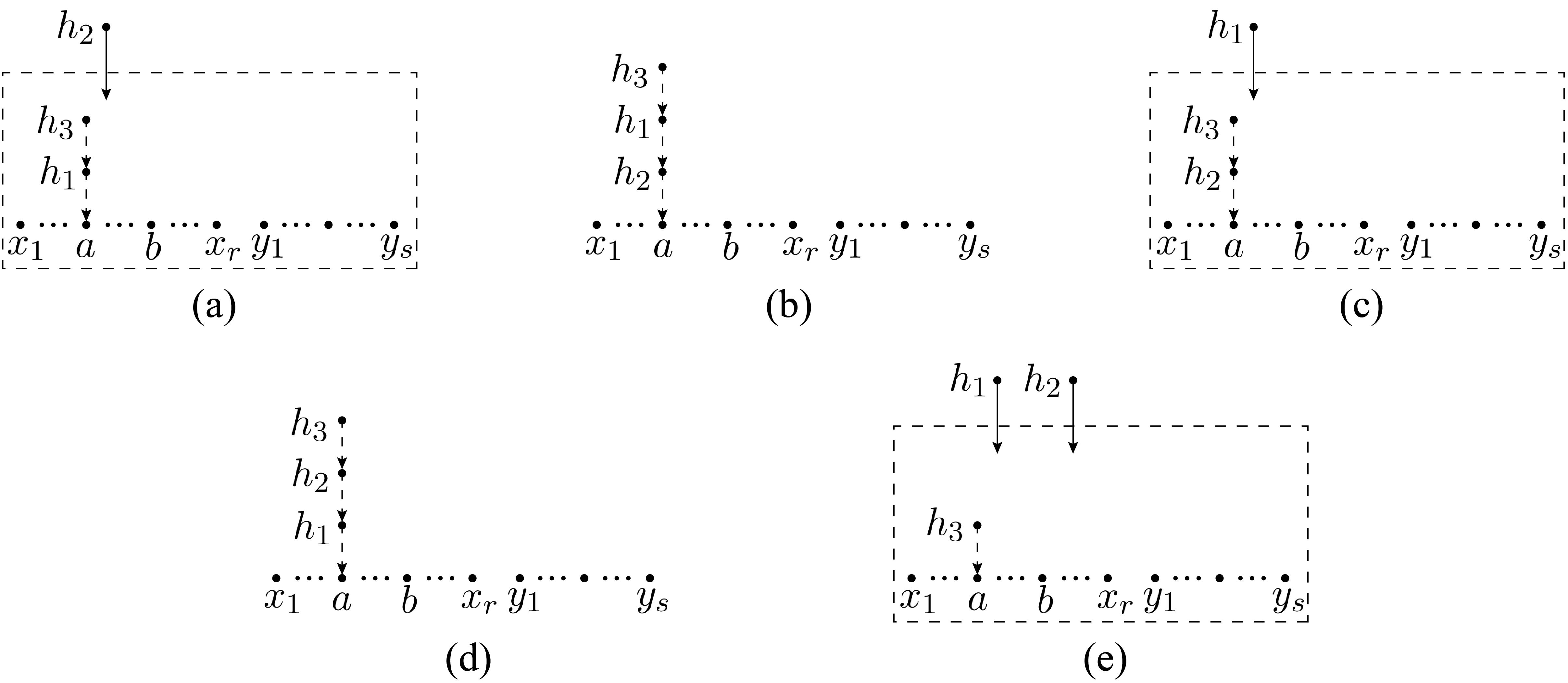}
\caption{All three-graviton graphs with  $h_3$ as starting node of a chain pointing towards gluon $a$ via one or several type-4 edges}
\label{Fig:13}
\end{figure}

{\bf Part-1}: the part involving all terms  with $\Spbb{h_i,b}$ $(i=1,...,t)$.~~~~Such a term is given by the Pfaffian of the matrix $[\W{\Psi}]_{\mathsf{H},a,b;\mathsf{H}}$ when the two rows and two columns with respect to $h_i$ and $b$ are deleted. Since the new matrix has the form (\ref{Eq:Pfaffianexpansion0}), according to  \eqref{Eq:Pfaffianexpansion1}, the expansion coefficient of $\Spbb{h_i,b}$ reduces into the determinant of a $t\times t$ matrix, as follows
\bea
\begin{vmatrix}
\phi_{h_1} & \cdots & \cancel{-\frac{\Spaa{\xi,h_i}\Spbb{h_i,h_1}}{\Spaa{\xi,h_1}\Spaa{h_i,h_1}}}  & \cdots & -\frac{\Spaa{\xi,h_t}\Spbb{h_t,h_1}}{\Spaa{\xi,h_1}\Spaa{h_t,h_1}} & -\frac{\Spaa{\xi,a}\Spbb{a,h_1}}{\Spaa{\xi,h_1}\Spaa{a,h_1}}      \\
\vdots & \ddots & \vdots & \ddots & \vdots & \vdots    \\
\cancel{-\frac{\Spaa{\xi,h_1}\Spbb{h_1,h_i}}{\Spaa{\xi,h_i}\Spaa{h_1,h_i}}} & \cdots &\cancel{~~~\phi_{h_i}~~~} & \cdots & \cancel{-\frac{\Spaa{\xi,h_t}\Spbb{h_t,h_i}}{\Spaa{\xi,h_i}\Spaa{h_t,h_i}}} & \cancel{-\frac{\Spaa{\xi,a}\Spbb{a,h_i}}{\Spaa{\xi,h_i}\Spaa{a,h_i}}}     \\
\vdots & \ddots & \vdots & \ddots & \vdots & \vdots    \\
-\frac{\Spaa{\xi,h_1}\Spbb{h_1,h_t}}{\Spaa{\xi,h_t}\Spaa{h_1,h_t}} & \cdots  &\cancel{-\frac{\Spaa{\xi,h_i}\Spbb{h_i,h_t}}{\Spaa{\xi,h_t}\Spaa{h_i,h_t}}}& \cdots & \phi_{h_t} & -\frac{\Spaa{\xi,a}\Spbb{a,h_t}}{\Spaa{\xi,h_t}\Spaa{a,h_t}}  \\
-\frac{\Spaa{\xi,h_1}\Spbb{h_1,h_i}}{\Spaa{\xi,h_i}\Spaa{h_1,h_i}} & \cdots &\cancel{~~~\phi_{h_i}~~~}& \cdots & -\frac{\Spaa{\xi,h_t}\Spbb{h_t,h_i}}{\Spaa{\xi,h_i}\Spaa{h_t,h_i}} & -\frac{\Spaa{\xi,a}\Spbb{a,h_i}}{\Spaa{\xi,h_i}\Spaa{a,h_i}}
\end{vmatrix}\,.\Label{Eq:Determinant}
\eea
To show the expansion of this determinant, let us have a look at the $t=3$, $i=3$ example. This determinant can be expanded by the third row
\bea
\begin{vmatrix}
\phi_{h_1} &  -\frac{\Spaa{\xi,h_2}\Spbb{h_2,h_1}}{\Spaa{\xi,h_1}\Spaa{h_2,h_1}} & -\frac{\Spaa{\xi,a}\Spbb{a,h_1}}{\Spaa{\xi,h_1}\Spaa{a,h_1}}      \\
-\frac{\Spaa{\xi,h_1}\Spbb{h_1,h_2}}{\Spaa{\xi,h_2}\Spaa{h_1,h_2}} & \phi_{h_2} & -\frac{\Spaa{\xi,a}\Spbb{a,h_2}}{\Spaa{\xi,h_2}\Spaa{a,h_2}}  \\
-\frac{\Spaa{\xi,h_1}\Spbb{h_1,h_3}}{\Spaa{\xi,h_3}\Spaa{h_1,h_3}} & -\frac{\Spaa{\xi,h_2}\Spbb{h_2,h_3}}{\Spaa{\xi,h_3}\Spaa{h_2,h_3}} & -\frac{\Spaa{\xi,a}\Spbb{a,h_3}}{\Spaa{\xi,h_3}\Spaa{a,h_3}}
\end{vmatrix}&=&~\frac{\Spaa{\xi,h_1}\Spbb{h_1,h_3}}{\Spaa{\xi,h_3}\Spaa{h_1,h_3}}
\begin{vmatrix}
\phi_{h_2} & -\frac{\Spaa{\xi,a}\Spbb{a,h_2}}{\Spaa{\xi,h_2}\Spaa{a,h_2}}  \\
-\frac{\Spaa{\xi,h_2}\Spbb{h_2,h_1}}{\Spaa{\xi,h_1}\Spaa{h_2,h_1}} & -\frac{\Spaa{\xi,a}\Spbb{a,h_1}}{\Spaa{\xi,h_1}\Spaa{a,h_1}}
\end{vmatrix}\nn
&&~~~~+\frac{\Spaa{\xi,h_2}\Spbb{h_2,h_3}}{\Spaa{\xi,h_3}\Spaa{h_2,h_3}}
\begin{vmatrix}
\phi_{h_1} & -\frac{\Spaa{\xi,a}\Spbb{a,h_1}}{\Spaa{\xi,h_1}\Spaa{a,h_1}}  \\
-\frac{\Spaa{\xi,h_1}\Spbb{h_1,h_2}}{\Spaa{\xi,h_2}\Spaa{h_1,h_2}} & -\frac{\Spaa{\xi,a}\Spbb{a,h_2}}{\Spaa{\xi,h_2}\Spaa{a,h_2}}
\end{vmatrix}\nn
&&~~~~~~~~-\frac{\Spaa{\xi,a}\Spbb{a,h_3}}{\Spaa{\xi,h_3}\Spaa{a,h_3}}
\begin{vmatrix}
\phi_{h_1} &  -\frac{\Spbb{h_2,h_1}}{\Spaa{h_2,h_1}}  \\
-\frac{\Spbb{h_1,h_2}}{\Spaa{h_1,h_2}} & \phi_{h_2}
\end{vmatrix},
\eea
\begin{figure}
\centering
\includegraphics[width=0.85\textwidth]{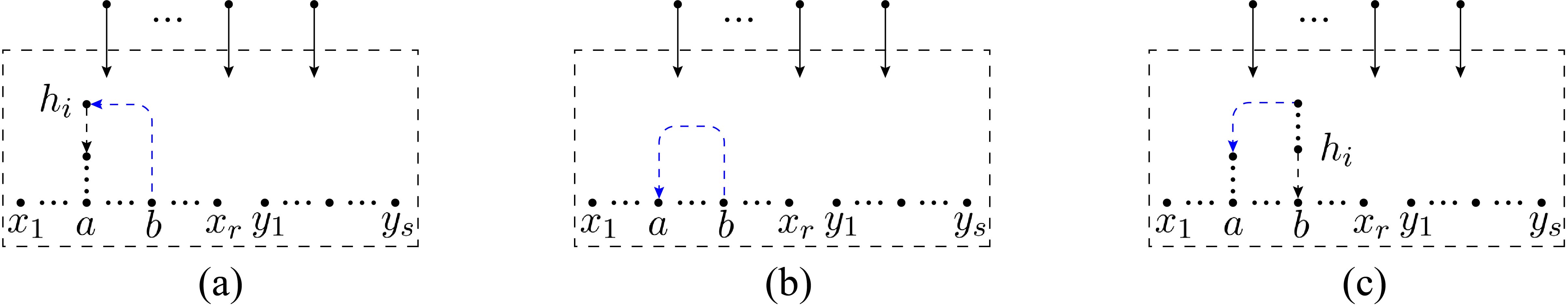}
\caption{Three typical graphs corresponding to the three parts, where (a): $h_i$ is directly connected to gluon $b$ via a type-3 edge $e(b,h_i)$, (b): gluons $a$ and $b$ are connected, and (c): $h_i$ is directly connected to gluon $b$ via a type-4 edge $e(h_i,b)$.}
\label{Fig:16}
\end{figure}
where the two second-order determinants in the first two terms have the same pattern (\ref{Eq:Determinant}), while the last term involves a Hodges determinant. Substituting the corresponding diagram \figref{Fig:15a} (a) for the second-order determinant into the first two terms respectively, one reduces the first two terms into  (a), (b) and (c), (d) in \figref{Fig:13}. The Hodges determinant can be expanded into the sum of all possible diagrams where all gluons play as roots and both gravitons $h_1$ and $h_2$ are connected to the roots via spanning forests. Thus the third term provides the graph \figref{Fig:13} (e). In general, we expand the determinant (\ref{Eq:Determinant}) by the last row. \emph{The contribution of all terms in this expansion provides all graphs of the structure \figref{Fig:16} (a), where $h_i$ is always the starting node of a chain pointing towards the gluon $a$ via type-4 edges, and is  directly connected to $b$ via a type-3 edge. Thus this determinant produces those graphs where all gravitons on the bridge between $a$ and $b$ are on the node $a$ side.}

{\bf Part-2}: the part containing a factor $\Spbb{a,b}$ and a Hodges determinant for all gravitons.~~~~According to the spanning forest of Hodges determinant, \emph{this part contributes \figref{Fig:16} (b) where $a$ and $b$ are directly connected to each other by a type-3 edge.}

{\bf Part-3}:  the part that is given by a factor $\frac{\Spaa{\xi,b}\Spbb{b,h_i}}{\Spaa{\xi,h_i}\Spaa{b,h_i}}$ multiplying to the following Pfaffian
\bea
&\text{Pf}&\,
\left[\begin{array}{ccccc|ccc}
0 & \cdots & \Spbb{h_t,h_1} & \Spbb{a,h_1} & \Spbb{h_i,h_1} & -\phi_{h_1} & \cdots & \frac{\Spaa{\xi,h_1}\Spbb{h_1,h_t}}{\Spaa{\xi,h_t}\Spaa{h_1,h_t}}      \\
\vdots & \ddots & \vdots & \vdots & \vdots & \vdots & \ddots & \vdots   \\
\Spbb{h_1,h_t} & \cdots & 0 & \Spbb{a,h_t} & \Spbb{h_i,h_t} & \frac{\Spaa{\xi,h_t}\Spbb{h_t,h_1}}{\Spaa{\xi,h_1}\Spaa{h_t,h_1}} & \cdots & -\phi_{h_t}   \\
\Spbb{h_1,a} & \cdots & \Spbb{h_t,a} & 0 & \Spbb{h_i,a} & \frac{\Spaa{\xi,a}\Spbb{a,h_1}}{\Spaa{\xi,h_1}\Spaa{a,h_1}} & \cdots & \frac{\Spaa{\xi,a}\Spbb{a,h_t}}{\Spaa{\xi,h_t}\Spaa{a,h_t}}   \\
\Spbb{h_1,h_i} & \cdots & \Spbb{h_t,h_i} & \Spbb{a,h_i} & 0 & \frac{\Spaa{\xi,h_i}\Spbb{h_i,h_1}}{\Spaa{\xi,h_1}\Spaa{h_i,h_1}} & \cdots & \frac{\Spaa{\xi,h_i}\Spbb{h_i,h_t}}{\Spaa{\xi,h_t}\Spaa{h_i,h_t}}   \\ \hline
\phi_{h_1} & \cdots & \frac{\Spaa{\xi,h_t}\Spbb{h_t,h_1}}{\Spaa{\xi,h_1}\Spaa{h_t,h_1}} &  -\frac{\Spaa{\xi,a}\Spbb{a,h_1}}{\Spaa{\xi,h_1}\Spaa{a,h_1}}  & -\frac{\Spaa{\xi,h_i}\Spbb{h_i,h_1}}{\Spaa{\xi,h_1}\Spaa{h_i,h_1}} & 0 & \cdots & 0   \\
\vdots & \ddots & \vdots & \vdots & \vdots & \vdots & \ddots & \vdots   \\
-\frac{\Spaa{\xi,h_1}\Spbb{h_1,h_t}}{\Spaa{\xi,h_t}\Spaa{h_1,h_t}} & \cdots & \phi_{h_t}  &  -\frac{\Spaa{\xi,a}\Spbb{a,h_t}}{\Spaa{\xi,h_t}\Spaa{a,h_t}}  & -\frac{\Spaa{\xi,h_i}\Spbb{h_i,h_t}}{\Spaa{\xi,h_t}\Spaa{h_i,h_t}} & 0 & \cdots & 0
\end{array}\right],
\eea
where the $i$-th row and  column have been adjusted to the $(t+1)$-th row and column.~~~~The above Pfaffian can be obtained from the Pfaffian $\text{Pf}\,[\W{\Psi}]_{\mathsf{H}',a,b;\mathsf{H}'}$ with $t-1$ gravitons in $\mathsf{H}'=\{h_1,...,h_{t-1}\}$, through the replacements $h_{i}\to h_{i+1}$, $h_{i+1}\to h_{i+2}$, $\cdots$, $h_{t-1}\to h_{t}$ and $b\to h_{i}$. Based on the graphs provided by Pfaffian $\text{Pf}\,[\W{\Psi}]_{\mathsf{H}',a,b;\mathsf{H}'}$ with $t-1$ gravitons, \emph{this part provides the \figref{Fig:16} (c), in which there is at least a graviton $h_i$ living on the $b$ side of the bridge between $a$, $b$, and $h_i$ is adjacent to $b$. The edge between $h_i$ and $b$ corresponds to the factor  $\frac{\Spaa{\xi,b}\Spbb{b,h_i}}{\Spaa{\xi,h_i}\Spaa{b,h_i}}$.}

To sum up, all the three parts of graphs together reproduce all graphs (with all possible configurations of the bridge between $a$ and $b$) that support \eqref{Eq:DoubleTraceMHV2}.

\section{Vanishing configurations}\label{SE:VanishingCoufigurations}
So far, we have calculated the double-trace MHV amplitudes with two negative-helicity gluons  by the CHY formula and proven that this result is equivalent to the symmetric one (\ref{Eq:DoubleTraceMHV1}). As pointed in \cite{Cachazo:2014xea,Tian:2021dzf}, the $m$ ($m\geq 3$)-trace amplitudes with two negative-helicity particles as well as the double-trace $(-,-)$-amplitudes with at least one negative-helicity graviton have to vanish. This fact, in the framework of CHY formula (\ref{Eq:IntegratedCHY}) in four dimensions, can be understood as follows: Neither the MHV solution nor any other solution supports these amplitudes. As already pointed in \cite{Weinzierl:2014vwa,Du:2016fwe},  the reduced Pfaffian, i.e. $\mathcal{I}_R$, in \eqref{Eq:EYMIntegrand} with the MHV configuration (i.e. the $(-,-)$-configuration) can only get nonvanishing contribution from the MHV solution. Therefore, one only needs to prove that the corresponding $\mathcal{I}_L$ with more than three traces or the double-trace one with at least one negative-helicity graviton has to vanish when the MHV solution (\ref{Eq:MHVSolution}) is substituted.   In the following, we first understand this point straightforwardly by using the MHV solution, then  provide a more generic discussion on the vanishing condition of multi-trace EYM amplitudes with an arbitrary total number of negative-helicity particles.

\subsection{The vanishing configurations with two negative-helicity particles}

One can make the following replacement of the double-trace MHV amplitude with $(g^-_i,g^-_j)$-configuration to  get the double-trace amplitude with $(g^-_i,h^-_1)$-configuration, in the MHV sector (i.e. the part supported by the MHV solution) of the Pfaffian in $\mathcal{I}_L$:
\bea
\frac{\epsilon^+_{h_1}(\xi)\cdot k_c}{\sigma_{h_1}-\sigma_c}=\frac{\Spaa{\xi,c}\Spbb{c,h_1}\Spaa{\zeta,h_1}\Spaa{\zeta,c}F}{\Spaa{\xi,h_1}\Spaa{h_1,c}}&\to& \frac{\epsilon^-_{h_1}(q)\cdot k_c}{\sigma_{h_1}-\sigma_c}=\frac{\Spbb{c,q}\Spaa{\zeta,h_1}\Spaa{\zeta,c}F}{\Spbb{q,h_1}}\\
-\Sl^n_{\substack{c=1\\c\ne {h_1}}}\frac{\epsilon^+_{h_1}(\xi)\cdot k_c}{\sigma_{h_1}-\sigma_c}\to -\Sl^n_{\substack{c=1\\c\ne {h_1}}}\frac{\epsilon^-_{h_1}(q)\cdot k_c}{\sigma_{h_1}-\sigma_c}=&-&\Sl^n_{\substack{c=1\\c\ne {h_1}}}\frac{\Spbb{c,q}\Spaa{\zeta,h_1}\Spaa{\zeta,c}F}{\Spbb{q,h_1}}=\frac{\Spbb{h_1,q}\Spaa{\zeta,h_1}\Spaa{\zeta,h_1}F}{\Spbb{q,h_1}}\\
\frac{\epsilon^+_{h_1}(\xi)\cdot \epsilon^+_{c}(\xi)}{\sigma_{h_1}-\sigma_c}=0&\to&\frac{\epsilon^-_{h_1}(q)\cdot \epsilon^+_{c}(\xi)}{\sigma_{h_1}-\sigma_c}
\eea
where $q$ is the reference momentum of negative-helicity graviton $h_1$, and $c$ can be any graviton or gluon except $h_1$. When the factors  $(\Spaa{h_1,\zeta}\dots\Spaa{h_t,\zeta})^2\Spaa{a,\zeta}\Spaa{b,\zeta}F^{2t+2}[qh_1]^{-1}$, $(\Spaa{h_1,\zeta}\dots\Spaa{h_t,\zeta})^2\Spaa{a,\zeta}\Spaa{b,\zeta}[qh_1]^{-1}$  are extracted out from the rows and columns respectively, the $\W{A}'$-block of the simplified matrix $\W{\Psi}'$ (see (\ref{Eq:ABCtilde})) is a $(t+3)\times(t+3)$ matrix
\bea
\W{A}'_{ab}=\Bigg\{
            \begin{array}{cc}
              \Spbb{b,a} &, a\ne b \\
               0 &,a=b \\
            \end{array}\,,\,\,\,\,a,b\in\{h_1,\dots,h_t,a,b,q\}\, ,
\eea
where, we have incorporated the row and column, whose entries have the form $\frac{\epsilon_{h_1}\cdot k_c}{\sigma_{h_1}-\sigma_c}$, corresponding to $h_1$ in the original $C$- and $-C^T$-blocks of the $\Psi$ into $\W{A}'$, respectively. \emph{ From this angle, the negative-helicity graviton $h_1$ seems like a gluon trace which contributes a pair of gluons $a$, $b$ with momenta $k_{h_1}$ and $q$, respectively.}  The element  $\frac{\epsilon_{h_1}\cdot \epsilon_{h_1}}{\sigma_{h_1}-\sigma_c}=0$, which comes from the original $B$-block is also incorporated, as the corresponding diagonal entry of $\W{A}'$. Correspondingly, the   $B$-block becomes a $t\times t$ matrix with an $(t-1)\times (t-1)$ vanishing block $\W{B}'$, and the $\W{C}'$-block is a $(t-1)\times(t+3)$ matrix.

Since the  $\W{A}'$-block has four rows more than $\W{C}'$-, $\W{B}'$-blocks, if we apply the recursive expansion (\ref{Eq:Pfaffianexpansion2}) repeatedly till all entries of the $\W{C}'$-block are extracted out of the Pfaffian, the original Pfaffian becomes a combination of the Pfaffians for $4\times 4$ skew matrices that all come from the $\W{A}'$-block. Specifically, such a Pfaffian has the following general form
\bea
\text{Pf}\,
\begin{bmatrix}
0 & \Spbb{d,c} & \Spbb{e,c} & \Spbb{f,c}      \\
\Spbb{c,d} & 0 & \Spbb{e,d} & \Spbb{f,d}     \\
\Spbb{c,e} & \Spbb{d,e} & 0 & \Spbb{f,e}      \\
\Spbb{c,f} & \Spbb{d,f}&\Spbb{e,f} & 0
\end{bmatrix}
=-(\Spbb{c,e}\Spbb{d,f}-\Spbb{d,e}\Spbb{c,f}-\Spbb{c,d}\Spbb{e,f})=0\,,
\Label{Eq:3.18}
\eea
which is zero due to Schouten identity. Thus, \emph{the double-trace MHV amplitude with $(g^-,h^-)$- configuration vanishes.}
Furthermore, if the helicity of more gravitons are chosen to be negative, more rows/columns coming from the  $B$- and $C$-blocks of the original Pfaffian will turn into rows and columns of $\W{A}'$-block, accompanied by the same momentum $q$. Therefore, the Pfaffian is finally expanded into combination of Pfaffians for $l\times l$ ($l>4$) submatrices of the $\W{A}'$-block. When further expanding these Pfaffians, one can always get a combination of Pfaffians with the pattern (\ref{Eq:3.18}), which has to vanish either due to Schouten identity or due to the vanishing of Pfaffian with two identical rows/columns. From this statement, \emph{we explicitly prove the fact that an $m$-trace ($m\geq 1$) amplitude with more than two negative-helicity gravitons are not supported by the MHV solution.}


Regardless of the helicity configuration, the $\W{A}$-block has at least 4 more rows/columns than the $\W{C}$-block for an $m$-trace $(m\geq 3)$ EYM amplitude. By recursively expanding the Pfaffian as shown in the above discussions, one finally gets a combination of Pfaffians with the pattern (\ref{Eq:3.18}). Therefore, \emph{the MHV solution to SE does not support the m-trace ($m\geq 3$) amplitude with any helicity configuration.}

The above analysis, together with the fact that $\cal{I}_R$ with two negative-helicity particles is only supported by the MHV solution, implies an amplitude with $(-,-)$-configuration has to vanish when it is (i). a single-trace amplitude with two negative-helicity gravitons, (ii). a double-trace amplitude with more than one negative-helicity graviton or (iii). an $m$-trace ($m\geq 3$) amplitude. All these three cases can be unified into the condition $n_g^-\leq m-1$ where $n_g^-$ is the number of negative-helicity gluons. This fact was also observed earlier in pure-gluon multi-trace case \cite{Cachazo:2014xea}. In the next subsection, we will see this condition can be generalized to amplitudes with an arbitrary total number of negative-helicity particles.

\subsection{The vanishing configurations with an arbitrary number of negative-helicity particles}
To study the vanishing configurations, one may try to follow the discussions in  \cite{Du:2016fwe}, which provides a discriminant matrix $\mathfrak{C}_{-}$ whose rank relies on solutions to SE. We find that the main results in \cite{Du:2016fwe} cannot be trivially extended to multi-trace case, because the proof of the crucial result (5.9) in \cite{Du:2016fwe} holds when the rank of the discriminant matrix $\mathfrak{C}_{-}$ satisfies   $\text{rank}(\mathfrak{C}_{-})\leq t^-$ ($t^-$ is the number of negative-helicity gravitons).  This is sufficient for single-trace discussions, however, in multi-trace case, more vanishing amplitudes exist: even all negative-helicity particles are gluons, the amplitude may also vanish, but the rank of $\mathfrak{C}_{-}$ is apparently larger than $t^-=0$. On another side,  following the line \cite{Witten:2003nn} and \cite{Cachazo:2013zc}, the work \cite{Cachazo:2014xea} has discussed the vanishing condition of multi-trace pure-gluon amplitudes and showed that in this case, the vanishing condition is that the number of negative-helicity gluons is less than the number of traces.  In this subsection, we combine the two approaches with the MHV example in the previous subsection to prove that an amplitude with any total number of negative-helicity particles (gravitons and gluons) $n^-$  vanishes when:
\bea
n_g^-\leq m-1. \Label{Eq:VanishingCondition}
\eea
Note that the number of negative-helicity particles $n^-$ is always assumed to be less than the number of positive-helicity ones $n^+$, i.e. $n^-<n^+$. If not, one can always exchange the roles between the positive-helicity particles and negative-helicity particles and then get the vanishing condition:  $n_g^+\leq m-1$.   The condition (\ref{Eq:VanishingCondition}) can be proved by combining the results proposed in \cite{Du:2016fwe} with the observation (\ref{Eq:3.18}) in the previous subsection. Now we review two critical points of  \cite{Du:2016fwe}:
\begin{itemize}
\item (i). As pointed in  \cite{Du:2016fwe}, solution set to SE can be given by the union of disjoint subsets ${\sf{P}}_{-}(n-3,l)$ ($l=0,...,n-4$).  Each subset ${\sf{P}}_{-}(n-3,l)$ is the collection of solutions such that the rank of $n\times n$ matrix\footnote{This matrix is closely related to the $\Phi$ in \cite{Cachazo:2012kg,Cachazo:2012pz} which provided a twistor string approach to the integrand for $N=8$ supergravity. }
\bea
(\mathfrak{C}_{-})_{ab}&=&\Biggl\{\begin{array}{cc}\frac{\Spaa{ab}}{\sigma_{ab}}& ,a\neq b\\ -\sum_{\scriptsize\substack{c=1\\c\neq a}}^{n}\frac{\Spaa{ac}\Spbb{cq}}{\sigma_{ac}\Spbb{aq}}&,a=b \end{array}\Label{Eq:DiscriminantMatrix}
\eea
is $r=\text{rank}(\mathfrak{C}_{-})=\text{rank}({C}_{-})=l+1$. For example, the solution  ${\sf{P}}_{-}(n-3,0)$ (in fact the MHV solution (\ref{Eq:MHVSolution})) makes the $\mathfrak{C}_{-}$ of rank $1$.

\item (ii). As shown in  \cite{Du:2016fwe}, the reduced Pfaffian $\mathcal{I}_R$ in (\ref{Eq:EYMIntegrand}) with $n^-$ negative-helicity particles is only supported by solutions in  ${\sf{P}}_{-}(n-3,l)$ where $n^-=l+2$. For instance, when $n^-=2$, $l$ can only be $0$, corresponding to the MHV solution (\ref{Eq:MHVSolution}).

\end{itemize}
With the above facts in hands, let us study the support of solutions on the  $\mathcal{I}_L$ in (\ref{Eq:EYMIntegrand}), or more concretely on the Pfaffian
\bea
\text{Pf}\,[\Psi]_{\mathsf{H},a_1,b_1,\dots,a_{m-1},b_{m-1};\mathsf{H}}=\text{Pf}\,\left[\begin{array}{c|c|c}A_{(t+2m-2)\times(t+2m-2)}&-(C_{-}^T)_{(t+2m-2)\times t^-}  &-(C_{+}^T)_{(t+2m-2)\times t^+} \\ \hline\,
(C_{-})_{t^-\times(t+2m-2)} & O_{t^-\times t^-} & -(B^T)_{t^-\times t^+} \\ \hline
(C_{+})_{t^+\times(t+2m-2)} & B_{t^+\times t^-} & O_{t^+\times t^+}\\
\end{array}\right].\Label{Eq:Pf1}
\eea
We first extract out all entries of the $B$-block and the $C_+$-block, using (\ref{Eq:Pfaffianexpansion2}) repeatedly. The Pfaffian (\ref{Eq:Pf1}) is thus expanded in terms of Pfaffians with the following form
\bea
\text{Pf}\,\left[\begin{array}{c|c}A_{(t^-+t'+2m-2)\times(t^-+t'+2m-2)}&-(C_{-}^T)_{(t^-+t'+2m-2)\times (t^--t')} \\ \hline\,
(C_{-})_{(t^--t')\times(t^-+t'+2m-2)} & O_{(t^--t')\times (t^--t')}
\end{array}\right],\Label{Eq:Pf2}
\eea
where $t'$ ($0\leq t'\leq t^-$) is the number of rows we delete in the original $C_{-}$-block, and the row and column indices of $A_{(t^-+t'+2m-2)\times(t^-+t'+2m-2)}$  become elements of an order-$(t^-+t'+2m-2)$ subset of $\mathsf{H}\cup\{a_1,b_1\}\cup...\cup\{a_{m-1},b_{m-1}\} $. Noting that the entries of $A$ and $C_{-}$ can be expressed by those in the discriminant matrix (\ref{Eq:DiscriminantMatrix}):
\bea
A_{ij}&=&(\mathfrak{C}_{-})_{ij}\Spbb{i,j},~~~~~(C^-)_{ij}=(\mathfrak{C}_{-})_{ij}\frac{\Spbb{j,q}}{\Spbb{i,q}},
\eea
we rewrite the Pfaffian (\ref{Eq:Pf2}) as
\bea
{\footnotesize\left(\prod\limits_{p\in \mathsf{H}'}\frac{1}{\Spbb{p,q}}\right)\text{Pf}\,X\equiv\left(\prod\limits_{p\in \mathsf{H}'}\frac{1}{\Spbb{p,q}}\right)\text{Pf}\,{\scriptsize\left[\begin{array}{c|c}\big((\mathfrak{C}_{-})_{ij}\Spbb{i,j}\big)_{(t^-+t'+2m-2)\times(t^-+t'+2m-2)}&-{\big((\mathfrak{C}^T_{-})_{ii'}{\Spbb{i,q}}\big)_{(t^-+t'+2m-2)\times (t^--t')}}  \\ \hline\,
{\big((\mathfrak{C}_{-})_{i'j}{\Spbb{j,q}}\big)_{(t^--t')\times(t^-+t'+2m-2)}} & O_{(t^--t')\times (t^--t')}
\end{array}\right]}.~~~\Label{Eq:Pf3}}\nn
\eea
Here, $\mathsf{H}'$ denote the $(t^--t')$-subset of the $\mathsf{H}^-$  and the common numerators of rows/columns in ${C}_{-}$ have been extracted out. For a given solution of SE in ${\sf{P}}_{-}(n-3,l)$, the rank of $\mathfrak{C}_{-}$ is $r=l+1$, this means any column of $(\mathfrak{C}_{-})_{n\times n}$ can be expanded in terms of $r$ independent columns $\mathbf{c}^1$, $\mathbf{c}^2$,..., $\mathbf{c}^r$. Each column $\mathbf{c}^k$ ($k=1,...,r$) has the form
\bea
\mathbf{c}^k\equiv\left[\mathbf{c}^k_1,\mathbf{c}^k_2,...,\mathbf{c}^k_n\right]^T=\left[\left(\mathfrak{C}_{-}\right)_{1\,k},\left(\mathfrak{C}_{-}\right)_{2\,k},...,\left(\mathfrak{C}_{-}\right)_{n\,k}\right]^T,
\eea
where, without loss of generality, we have supposed that the first $r$ columns of $(\mathfrak{C}_{-})_{n\times n}$ are the independent columns.
Since the $(\mathfrak{C}_{-})_{n\times n}$ is a symmetric matix, columns of $(\mathfrak{C}_{-})^T_{n\times n}$ can also be expanded by  $\mathbf{c}^k$. When the $(\mathfrak{C}_{-})_{i\,j}$ in each column of $X$ is expanded as a combination of $\left(\mathfrak{C}_{-}\right)_{i\,k}$, the determinant of $X$ is finally given by a combination of determinants
\bea
{\footnotesize\text{\bf det}\,X\sim \text{Combination of}~\text{\bf det}\left[\begin{array}{c}{\mathbf{c}_{i_1}^{k_1}}\Spbb{i_1,i_1}\\ \mathbf{c}_{i_2}^{k_1}\Spbb{i_2,i_1}\\\vdots\\\mathbf{c}_{i_{n'}}^{k_1}\Spbb{i_{n'},i_1}\\{\mathbf{c}_{i'_1}^{k_1}\Spbb{i_1,q}}\\ \mathbf{c}_{i'_2}^{k_1}\Spbb{i_1,q}\\\vdots\\  \mathbf{c}_{i'_{n''}}^{k_1}\Spbb{i_1,q}\end{array}
\begin{array}{c}{\mathbf{c}_{i_1}^{k_2}}\Spbb{i_1,i_2}\\ \mathbf{c}_{i_2}^{k_2}\Spbb{i_2,i_2}\\\vdots\\\mathbf{c}_{i_{n'}}^{k_2}\Spbb{i_{n'},i_2}\\{\mathbf{c}_{i'_1}^{k_2}}\Spbb{i_2,q}\\ \mathbf{c}_{i'_2}^{k_2}\Spbb{i_2,q}\\\vdots\\\mathbf{c}_{i'_{n''}}^{k_2}\Spbb{i_2,q}\end{array}
\begin{array}{c}\dots\\\dots\\\ddots\\ \dots \\\dots\\\dots\\\ddots\\\dots\end{array}
\begin{array}{c}{\mathbf{c}_{i_1}^{k_{n'}}}\Spbb{i_1,i_{n'}}\\ \mathbf{c}_{i_2}^{k_{n'}}\Spbb{i_2,i_{n'}}\\\vdots\\\mathbf{c}_{i_{n'}}^{k_{n'}}\Spbb{i_{n'},i_{n'}}\\{\mathbf{c}_{i'_1}^{k_{n'}}}\Spbb{i_{n'},q}\\ \mathbf{c}_{i'_2}^{k_{n'}}\Spbb{i_{n'},q}\\\vdots\\\mathbf{c}_{i'_{n''}}^{k_{n'}}\Spbb{i_{n'},q}\end{array}
\begin{array}{c}{\mathbf{c}_{i_1}^{k'_1}}\Spbb{q,i_1}\\ \mathbf{c}_{i_2}^{k'_1}\Spbb{q,i_2}\\\vdots\\\mathbf{c}_{i_{n'}}^{k'_1}\Spbb{q,i_{n'}}\\0\\ 0\\\vdots\\0\end{array}
\begin{array}{c}{\mathbf{c}_{i_1}^{k'_2}}\Spbb{q,i_1}\\ \mathbf{c}_{i_2}^{k'_2}\Spbb{q,i_2}\\\vdots\\\mathbf{c}_{i_{n'}}^{k'_2}\Spbb{q,i_{n'}}\\0\\ 0\\\vdots\\0\end{array}
\begin{array}{c}\dots\\\dots\\\ddots\\ \dots \\\dots\\\dots\\\ddots\\\dots\end{array}
\begin{array}{c}{\mathbf{c}_{i_1}^{k'_{n''}}}\Spbb{q,i_{1}}\\ \mathbf{c}_{i_2}^{k'_{n''}}\Spbb{q,i_{2}}\\\vdots\\\mathbf{c}_{i_{n'}}^{k'_{n''}}\Spbb{q,i_{n'}}\\0\\ 0\\\vdots\\0\end{array}
\right].~~~\Label{Eq:DetX}}
\eea
Here, we always move the rows/columns corresponding to negative-helicity gravitons to the first $t^-$ rows/columns for convenience, $n'$ and $n''$ are respectively given by $n'=t^-+t'+2m-2$ and  $n''=t^--t'$. Each of the superscripts $k_1,...,k_{n'=t^-+t'+2m-2}$ or $k'_1,...,k'_{n''=t^--t'}$ can take any value of $1,2,...,r$.  If we require the full determinant $\text{\bf det}\,X$ (thus $\text{ Pf}\,X$) vanish, we just require all terms in the expansion (\ref{Eq:DetX}) to be zero. As proved in \appref{sec:DetX}, each determinant on the RHS of (\ref{Eq:DetX}) for all possible  $0\leq t'\leq t^-$ vanishes if there exist at least three columns with the same $k$. This condition is satisfied by all determinmants on the RHS of (\ref{Eq:DetX}) only when the number of columns $n'+n''$ satisfy $n'+n''=2t^-+2m-2> 2r$, according to the pigeonhole principle. Therefore we get the vanishing condition (\ref{Eq:VanishingCondition}) when the facts  $r+1=l+2=n^-$ and $n^-=t^-+n_g^-$ are used.


\section{Conclusions}\label{SE:Conclusions}
In this note, we studied the relation between the multi-trace EYM amplitudes and the sectors of CHY amplitudes that are supported by distinct solutions to SE. For the double-trace MHV amplitudes with two negative-helicity gravitons, the MHV sector of CHY formula was shown to be reduced into a spanning forest form which consists of two mutually disconnected components. The latter was proven to be equivalent to the symmetric form proposed in \cite{Tian:2021dzf}.  We then show that the MHV sector of the CHY formula for the double-trace amplitudes with $(g^-,h^-)$- or $(h^-,h^-)$-configurations and all multi-trace amplitudes has to vanish. Moreover, we further prove that an $m$-trace EYM amplitude where the number of negative- (and/or positive-) helicity gluons is less than the number of gluon traces has to vanish.  This work provides an analysis of the nonvanishing double-trace MHV amplitude and the vanishing configurations for EYM amplitudes. We expect an analysis of nonvanishing configurations beyond MHV by either the graphic expansion or CHY formula,  the observation that a graviton seems like a gluon trace may play  a role in the further work.

%
\section*{Acknowledgements}
This work is supported by NSFC under Grant No. 11875206.

\appendix

\section{One-graviton example for the proof in \secref{Sec:GenProof}}\label{app:one-graviton}
\begin{figure}
\centering
\includegraphics[width=0.79\textwidth]{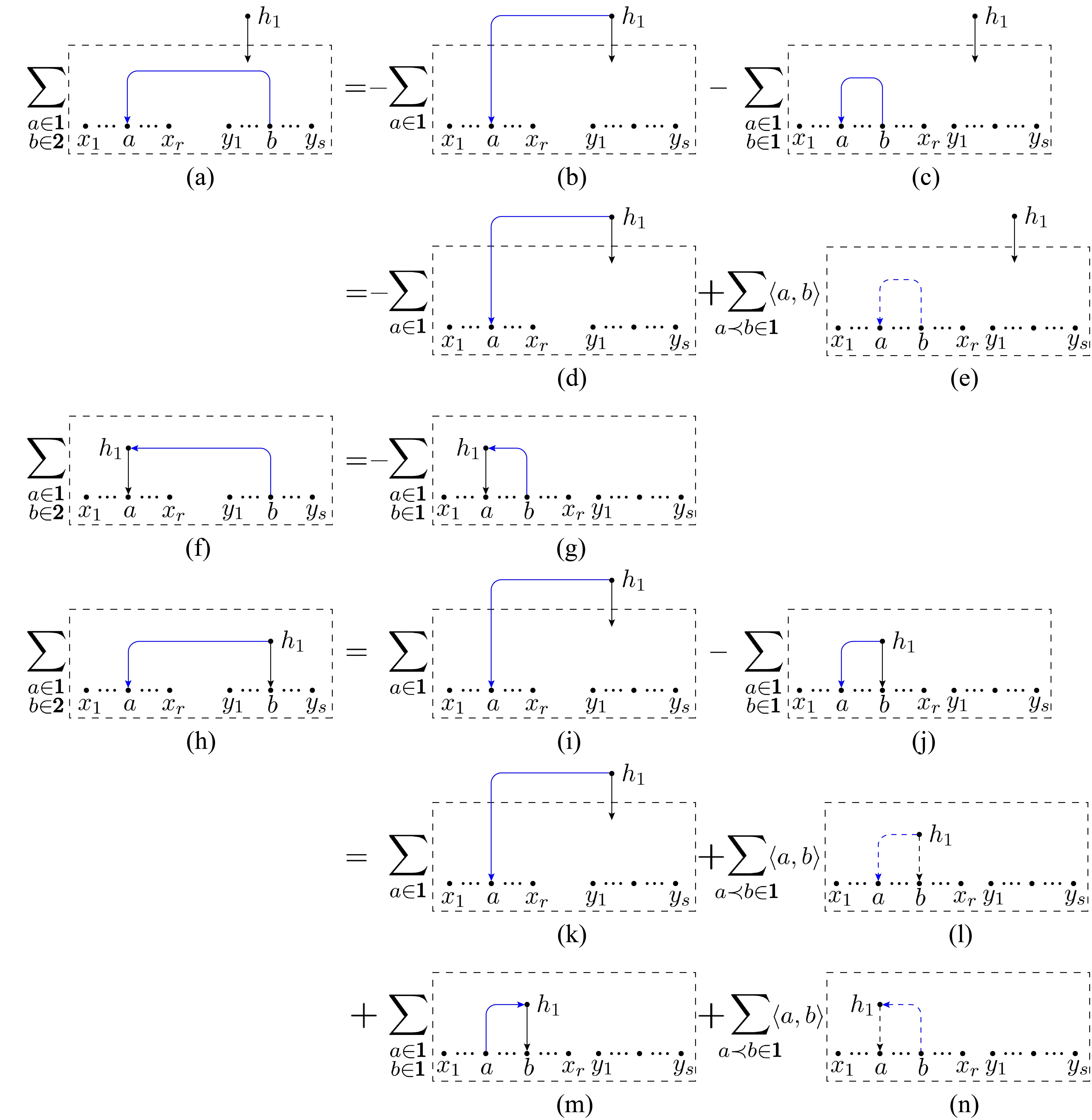}
\caption{Transformations of spanning forests for one-graviton $(g^-,g^-)$-amplitude}
\label{Fig:example2}
\end{figure}
As shown by \figref{Fig:example2}, there are three spanning forests (a), (f) and (h) for the amplitude with only one graviton $h_1$, which correspond to the three cases that the graviton is (i) outside the bridge, (ii).  on the LHS of the bridge  and (iii) on the RHS of the bridge, respectively.

After transforming spanning forest (a) with transformation-3 and then the resulted graph (c) with transformation-1, graphs (d) and (e) are obtained. Graph (g) also comes from the transformation-3 of spanning forest (f). For graph (h), we apply transformation-2 to get graphs (i) and (j), and then apply transformation-2 to graph (j) to get graphs (l), (m), and (n). Since graphs (d) and (k) (graphs (g) and (m)) are the same except for the signs, after they cancel with each other out, the spanning forests for one-graviton amplitude only provide graphs (e), (l) and (n) with bridges formed by type-3 and type-4 edges. Therefore, the formula (\ref{Eq:DoubleTraceMHV1}) and the formula (\ref{Eq:DoubleTraceMHV2}) for $(g^-,g^-)$-amplitude with one graviton are equivalent.

\section{The vanishing condition of $\text{\bf det}\,X$}\label{sec:DetX}

Now we consider a matrix on the RHS of (\ref{Eq:DetX}), in which three columns have the same  $k$. Any $3\times 3$ minor corresponding to these three columns has the form:
\bea
\text{\bf det}\left[\begin{array}{ccc}\mathbf{c}_{x_1}^{k} a_{x_1y_1} &\mathbf{c}_{x_1}^{k} {a}_{x_1y_2}  &\mathbf{c}_{x_1}^{k} a_{x_1y_3} \\\mathbf{c}_{x_2}^{k} a_{x_2y_1} &\mathbf{c}_{x_2}^{k} a_{x_2y_2}  &\mathbf{c}_{x_2}^{k} a_{x_2y_3} \\\mathbf{c}_{x_3}^{k} a_{x_3y_1} &\mathbf{c}_{x_3}^{k} a_{x_3y_2}  &\mathbf{c}_{x_3}^{k} a_{x_3y_3} \end{array}\right],\Label{Eq:DetX1}
\eea
where the form of $a_{xy}$ depends on its position in the matrix (\ref{Eq:DetX})
\bea
 a_{xy}\equiv\left\{\begin{array}{cc}\Spbb{x,y}&~~~(\text{if}~a_{xy}~\text{comes from the original matrix}~A )\\
 \Spbb{q,x}&~~~(\text{if}~a_{xy}~\text{comes from the original matrix}~-C_{-}^T )\\
 \Spbb{y,q} & ~~~(\text{if}~a_{xy}~\text{comes from the original matrix}~C)\\
 0&~~~(\text{if}~a_{xy}~\text{comes from the original matrix}~O )\end{array}\right. .\Label{Eq:axy}
\eea
Each row of determinant (\ref{Eq:DetX1}) has a common factor, $\mathbf{c}_{x_1}^{k}$, $\mathbf{c}_{x_2}^{k}$ or $\mathbf{c}_{x_3}^{k}$ that can be extracted out. Consequently, the determinant (\ref{Eq:DetX1}) is proportional to
\bea
\text{\bf det}\left[\begin{array}{ccc} a_{x_1y_1} & {a}_{x_1y_2}  &a_{x_1y_3} \\ a_{x_2y_1} & a_{x_2y_2}  & a_{x_2y_3} \\ a_{x_3y_1} & a_{x_3y_2}  & a_{x_3y_3} \end{array}\right].\Label{Eq:DetX2}
\eea
Now we show that the determinant (\ref{Eq:DetX2}) vanishes in all possible situations:
\begin{itemize}
\item If (\ref{Eq:DetX2}) involves at least two rows or two columns obtained from  $C_-$- or $-C_{-}^T$-blocks, these two columns or rows must be identical, because of (\ref{Eq:axy}). Thus the determinant vanishes.
\item If (\ref{Eq:DetX2}) invloves one row or one column obtained from  $C_-$- or $-C_{-}^T$-blocks, this column or row can be considered as a column or row which comes from the block $A$ while replacing the momentum of the particle by the reference momentum  $q$. This has been seen in the MHV case in the previous subsection. Thus we only need to consider the remaining case that all rows and columns correspond to gluons and/or gravitons that come from block $A$.
\item When all rows and columns refer to gluons and/or gravitons in original $A$-block, the determinant (\ref{Eq:DetX2}) in general has the pattern
\bea
\text{\bf det}\left[\begin{array}{ccc} \Spbb{x_1y_1} & \Spbb{x_1y_2}  &\Spbb{x_1y_3} \\ \Spbb{x_2y_1} & \Spbb{x_2y_2}  & \Spbb{x_2y_3} \\ \Spbb{x_3y_1} & \Spbb{x_3y_2}  & \Spbb{x_3y_3} \end{array}\right],\Label{Eq:DetX3}
\eea
which is further expanded as
\bea
&&\Spbb{x_1,y_1}\left(\Spbb{x_2,y_2}\Spbb{x_3,y_3}-\Spbb{x_2,y_3}\Spbb{x_3,y_2}\right)-\Spbb{x_1,y_2}\left(\Spbb{x_2,y_1}\Spbb{x_3,y_3}-\Spbb{x_2,y_3}\Spbb{x_3,y_1}\right)\nn
&+&\Spbb{x_1,y_3}\left(\Spbb{x_2,y_1}\Spbb{x_3,y_2}-\Spbb{x_2,y_2}\Spbb{x_3,y_1}\right).
\eea
By the use of Schouten identity, one can straightforwardly verify that the above expression vanishes for either (i). the row labels $x_1$, $x_2$, $x_3$ and  the column labels $y_1$, $y_2$, $y_3$ refer to distinct elements, or (ii). some of row indices are identical to some of the column indices (i.e. there exist zero entries which come from the diagonal entries of the original matrix $X$ ).
\end{itemize}

\bibliographystyle{JHEP}
\bibliography{NoteEYM4dRevison}

\providecommand{\href}[2]{#2}\begingroup\raggedright\begin{thebibliography}{10}

\bibitem{Stieberger:2016lng}
S.~Stieberger and T.~R. Taylor, {\it {New relations for Einstein-Yang-Mills
  amplitudes}},  {\em Nucl. Phys.} {\bf B913} (2016) 151--162,
  [\href{http://arxiv.org/abs/1606.09616}{{\tt arXiv:1606.09616}}].

\bibitem{Nandan:2016pya}
D.~Nandan, J.~Plefka, O.~Schlotterer, and C.~Wen, {\it {Einstein-Yang-Mills
  from pure Yang-Mills amplitudes}},  {\em JHEP} {\bf 10} (2016) 070,
  [\href{http://arxiv.org/abs/1607.05701}{{\tt arXiv:1607.05701}}].

\bibitem{Schlotterer:2016cxa}
O.~Schlotterer, {\it {Amplitude relations in heterotic string theory and
  Einstein-Yang-Mills}},  {\em JHEP} {\bf 11} (2016) 074,
  [\href{http://arxiv.org/abs/1608.00130}{{\tt arXiv:1608.00130}}].

\bibitem{Fu:2017uzt}
C.-H. Fu, Y.-J. Du, R.~Huang, and B.~Feng, {\it {Expansion of
  Einstein-Yang-Mills Amplitude}},  {\em JHEP} {\bf 09} (2017) 021,
  [\href{http://arxiv.org/abs/1702.08158}{{\tt arXiv:1702.08158}}].

\bibitem{Chiodaroli:2017ngp}
M.~Chiodaroli, M.~Gunaydin, H.~Johansson, and R.~Roiban, {\it {Explicit
  Formulae for Yang-Mills-Einstein Amplitudes from the Double Copy}},  {\em
  JHEP} {\bf 07} (2017) 002, [\href{http://arxiv.org/abs/1703.00421}{{\tt
  arXiv:1703.00421}}].

\bibitem{Teng:2017tbo}
F.~Teng and B.~Feng, {\it {Expanding Einstein-Yang-Mills by Yang-Mills in CHY
  frame}},  {\em JHEP} {\bf 05} (2017) 075,
  [\href{http://arxiv.org/abs/1703.01269}{{\tt arXiv:1703.01269}}].

\bibitem{Du:2017gnh}
Y.-J. Du, B.~Feng, and F.~Teng, {\it {Expansion of All Multitrace Tree Level
  EYM Amplitudes}},  \href{http://arxiv.org/abs/1708.04514}{{\tt
  arXiv:1708.04514}}.

\bibitem{Bern:2008qj}
Z.~Bern, J.~J.~M. Carrasco, and H.~Johansson, {\it {New Relations for
  Gauge-Theory Amplitudes}},  {\em Phys. Rev.} {\bf D78} (2008) 085011,
  [\href{http://arxiv.org/abs/0805.3993}{{\tt arXiv:0805.3993}}].

\bibitem{Du:2017kpo}
Y.-J. Du and F.~Teng, {\it {BCJ numerators from reduced Pfaffian}},  {\em JHEP}
  {\bf 04} (2017) 033, [\href{http://arxiv.org/abs/1703.05717}{{\tt
  arXiv:1703.05717}}].

\bibitem{Du:2018khm}
Y.-J. Du and Y.~Zhang, {\it {Gauge invariance induced relations and the
  equivalence between distinct approaches to NLSM amplitudes}},  {\em JHEP}
  {\bf 07} (2018) 177, [\href{http://arxiv.org/abs/1803.01701}{{\tt
  arXiv:1803.01701}}].

\bibitem{Hou:2018bwm}
L.~Hou and Y.-J. Du, {\it {A graphic approach to gauge invariance induced
  identity}},  {\em JHEP} {\bf 05} (2019) 012,
  [\href{http://arxiv.org/abs/1811.12653}{{\tt arXiv:1811.12653}}].

\bibitem{Du:2019vzf}
Y.-J. Du and L.~Hou, {\it {A graphic approach to identities induced from
  multi-trace Einstein-Yang-Mills amplitudes}},  {\em JHEP} {\bf 05} (2020)
  008, [\href{http://arxiv.org/abs/1910.04014}{{\tt arXiv:1910.04014}}].

\bibitem{Wu:2021exa}
K.~Wu and Y.-J. Du, {\it {Off-shell extended graphic rule and the expansion of
  Berends-Giele currents in Yang-Mills theory}},  {\em JHEP} {\bf 01} (2022)
  162, [\href{http://arxiv.org/abs/2109.14462}{{\tt arXiv:2109.14462}}].

\bibitem{Du:2022vsw}
Y.-J. Du and K.~Wu, {\it {Note on graph-based BCJ relation for Berends-Giele
  currents}},  \href{http://arxiv.org/abs/2207.02374}{{\tt arXiv:2207.02374}}.

\bibitem{Tian:2021dzf}
H.~Tian, E.~Gong, C.~Xie, and Y.-J. Du, {\it {Evaluating EYM amplitudes in four
  dimensions by refined graphic expansion}},  {\em JHEP} {\bf 04} (2021) 150,
  [\href{http://arxiv.org/abs/2101.02962}{{\tt arXiv:2101.02962}}].

\bibitem{Porkert:2022efy}
F.~Porkert and O.~Schlotterer, {\it {One-loop amplitudes in Einstein-Yang-Mills
  from forward limits}},  \href{http://arxiv.org/abs/2201.12072}{{\tt
  arXiv:2201.12072}}.

\bibitem{Zhou:2022djx}
K.~Zhou, {\it {Transmutation operators and expansions for $1$-loop Feynman
  integrands}},  \href{http://arxiv.org/abs/2201.01552}{{\tt
  arXiv:2201.01552}}.

\bibitem{Faller:2018vdz}
J.~Faller and J.~Plefka, {\it {Positive helicity Einstein-Yang-Mills amplitudes
  from the double copy method}},  {\em Phys. Rev. D} {\bf 99} (2019), no.~4
  046008, [\href{http://arxiv.org/abs/1812.04053}{{\tt arXiv:1812.04053}}].

\bibitem{Parke:1986gb}
S.~J. Parke and T.~R. Taylor, {\it {An Amplitude for $n$ Gluon Scattering}},
  {\em Phys. Rev. Lett.} {\bf 56} (1986) 2459.

\bibitem{Nguyen:2009jk}
D.~Nguyen, M.~Spradlin, A.~Volovich, and C.~Wen, {\it {The Tree Formula for MHV
  Graviton Amplitudes}},  {\em JHEP} {\bf 07} (2010) 045,
  [\href{http://arxiv.org/abs/0907.2276}{{\tt arXiv:0907.2276}}].

\bibitem{Hodges:2012ym}
A.~Hodges, {\it {A simple formula for gravitational MHV amplitudes}},
  \href{http://arxiv.org/abs/1204.1930}{{\tt arXiv:1204.1930}}.

\bibitem{Feng:2012sy}
B.~Feng and S.~He, {\it {Graphs, determinants and gravity amplitudes}},  {\em
  JHEP} {\bf 10} (2012) 121, [\href{http://arxiv.org/abs/1207.3220}{{\tt
  arXiv:1207.3220}}].

\bibitem{Du:2016wkt}
Y.-J. Du, F.~Teng, and Y.-S. Wu, {\it {Direct Evaluation of $n$-point
  single-trace MHV amplitudes in 4d Einstein-Yang-Mills theory using the CHY
  Formalism}},  {\em JHEP} {\bf 09} (2016) 171,
  [\href{http://arxiv.org/abs/1608.00883}{{\tt arXiv:1608.00883}}].

\bibitem{Cachazo:2014nsa}
F.~Cachazo, S.~He, and E.~Y. Yuan, {\it {Einstein-Yang-Mills Scattering
  Amplitudes From Scattering Equations}},  {\em JHEP} {\bf 01} (2015) 121,
  [\href{http://arxiv.org/abs/1409.8256}{{\tt arXiv:1409.8256}}].

\bibitem{Cachazo:2013gna}
F.~Cachazo, S.~He, and E.~Y. Yuan, {\it {Scattering equations and
  Kawai-Lewellen-Tye orthogonality}},  {\em Phys. Rev.} {\bf D90} (2014), no.~6
  065001, [\href{http://arxiv.org/abs/1306.6575}{{\tt arXiv:1306.6575}}].

\bibitem{Cachazo:2013hca}
F.~Cachazo, S.~He, and E.~Y. Yuan, {\it {Scattering of Massless Particles in
  Arbitrary Dimensions}},  {\em Phys. Rev. Lett.} {\bf 113} (2014), no.~17
  171601, [\href{http://arxiv.org/abs/1307.2199}{{\tt arXiv:1307.2199}}].

\bibitem{Cachazo:2013iea}
F.~Cachazo, S.~He, and E.~Y. Yuan, {\it {Scattering of Massless Particles:
  Scalars, Gluons and Gravitons}},  {\em JHEP} {\bf 07} (2014) 033,
  [\href{http://arxiv.org/abs/1309.0885}{{\tt arXiv:1309.0885}}].

\bibitem{Cachazo:2014xea}
F.~Cachazo, S.~He, and E.~Y. Yuan, {\it {Scattering Equations and Matrices:
  From Einstein To Yang-Mills, DBI and NLSM}},  {\em JHEP} {\bf 07} (2015) 149,
  [\href{http://arxiv.org/abs/1412.3479}{{\tt arXiv:1412.3479}}].

\bibitem{Du:2016blz}
Y.-j. Du, F.~Teng, and Y.-s. Wu, {\it {CHY formula and MHV amplitudes}},  {\em
  JHEP} {\bf 05} (2016) 086, [\href{http://arxiv.org/abs/1603.08158}{{\tt
  arXiv:1603.08158}}].

\bibitem{Weinzierl:2014vwa}
S.~Weinzierl, {\it {On the solutions of the scattering equations}},  {\em JHEP}
  {\bf 04} (2014) 092, [\href{http://arxiv.org/abs/1402.2516}{{\tt
  arXiv:1402.2516}}].

\bibitem{Du:2016fwe}
Y.-J. Du, F.~Teng, and Y.-S. Wu, {\it {Characterizing the solutions to
  scattering equations that support tree-level N$^{k}$ MHV gauge/gravity
  amplitudes}},  {\em JHEP} {\bf 11} (2016) 088,
  [\href{http://arxiv.org/abs/1608.06040}{{\tt arXiv:1608.06040}}].

\bibitem{Roberts:1972abc}
D.~Roberts, {\em {Mathematical Structure of Dual Amplitudes}}.
\newblock PhD thesis, Durham University, 1972.
\newblock Available at Durham E-Theses online.

\bibitem{Fairlie:1972abc}
D.~Fairlie and D.~Roberts, ``{Dual Models without Tachyons - a New Approach}.''
  unpublished Durham preprint PRINT-72-2440, 1972.

\bibitem{Fairlie:2008dg}
D.~B. Fairlie, {\it {A Coding of Real Null Four-Momenta into World-Sheet
  Co-ordinates}},  {\em Adv. Math. Phys.} {\bf 2009} (2009) 284689,
  [\href{http://arxiv.org/abs/0805.2263}{{\tt arXiv:0805.2263}}].

\bibitem{Monteiro:2013rya}
R.~Monteiro and D.~O'Connell, {\it {The Kinematic Algebras from the Scattering
  Equations}},  {\em JHEP} {\bf 03} (2014) 110,
  [\href{http://arxiv.org/abs/1311.1151}{{\tt arXiv:1311.1151}}].

\bibitem{Geyer:2014fka}
Y.~Geyer, A.~E. Lipstein, and L.~J. Mason, {\it {Ambitwistor Strings in Four
  Dimensions}},  {\em Phys. Rev. Lett.} {\bf 113} (2014), no.~8 081602,
  [\href{http://arxiv.org/abs/1404.6219}{{\tt arXiv:1404.6219}}].

\bibitem{Mason:2013sva}
L.~Mason and D.~Skinner, {\it {Ambitwistor strings and the scattering
  equations}},  {\em JHEP} {\bf 07} (2014) 048,
  [\href{http://arxiv.org/abs/1311.2564}{{\tt arXiv:1311.2564}}].

\bibitem{Dolan:2015iln}
L.~Dolan and P.~Goddard, {\it {General Solution of the Scattering Equations}},
  \href{http://arxiv.org/abs/1511.09441}{{\tt arXiv:1511.09441}}.

\bibitem{Cardona:2015ouc}
C.~Cardona and C.~Kalousios, {\it {Elimination and recursions in the scattering
  equations}},  {\em Phys. Lett.} {\bf B756} (2016) 180--187,
  [\href{http://arxiv.org/abs/1511.05915}{{\tt arXiv:1511.05915}}].

\bibitem{Cachazo:2016sdc}
F.~Cachazo and G.~Zhang, {\it {Minimal Basis in Four Dimensions and Scalar
  Blocks}},  \href{http://arxiv.org/abs/1601.06305}{{\tt arXiv:1601.06305}}.

\bibitem{He:2016vfi}
S.~He, Z.~Liu, and J.-B. Wu, {\it {Scattering Equations, Twistor-string
  Formulas and Double-soft Limits in Four Dimensions}},  {\em JHEP} {\bf 07}
  (2016) 060, [\href{http://arxiv.org/abs/1604.02834}{{\tt arXiv:1604.02834}}].

\bibitem{Xu:1986xb}
Z.~Xu, D.-H. Zhang, and L.~Chang, {\it {Helicity Amplitudes for Multiple
  Bremsstrahlung in Massless Nonabelian Gauge Theories}},  {\em Nucl. Phys. B}
  {\bf 291} (1987) 392--428.

\bibitem{Witten:2003nn}
E.~Witten, {\it {Perturbative gauge theory as a string theory in twistor
  space}},  {\em Commun. Math. Phys.} {\bf 252} (2004) 189--258,
  [\href{http://arxiv.org/abs/hep-th/0312171}{{\tt hep-th/0312171}}].

\bibitem{Cachazo:2013zc}
F.~Cachazo, {\it {Resultants and Gravity Amplitudes}},
  \href{http://arxiv.org/abs/1301.3970}{{\tt arXiv:1301.3970}}.

\bibitem{Cachazo:2012kg}
F.~Cachazo and D.~Skinner, {\it {Gravity from Rational Curves in Twistor
  Space}},  {\em Phys. Rev. Lett.} {\bf 110} (2013), no.~16 161301,
  [\href{http://arxiv.org/abs/1207.0741}{{\tt arXiv:1207.0741}}].

\bibitem{Cachazo:2012pz}
F.~Cachazo, L.~Mason, and D.~Skinner, {\it {Gravity in Twistor Space and its
  Grassmannian Formulation}},  {\em SIGMA} {\bf 10} (2014) 051,
  [\href{http://arxiv.org/abs/1207.4712}{{\tt arXiv:1207.4712}}].

\end{thebibliography}\endgroup

\end{document}